\documentclass[aps,pra,showpacs,twocolumn]{revtex4}
\usepackage{graphics}
\usepackage{amsmath}
\usepackage{graphicx}
\usepackage{amsfonts}
\usepackage{amssymb}

\begin{document}
\title{Entanglement Reactivation in Separable Environments}

\begin{abstract}
Combining two entanglement-breaking channels into a
correlated-noise environment restores the distribution of
entanglement. Surprisingly, this reactivation can be induced by
the injection of separable correlations from the composite
environment. In any dimension (finite or infinite), we can
construct classically-correlated \textquotedblleft twirling\textquotedblright%
\ environments which are entanglement-breaking in the transmission of single
systems but entanglement-preserving when two systems are transmitted. Here
entanglement is simply preserved by the existence of decoherence-free
subspaces. Remarkably, even when such subspaces do not exist, a fraction of
the input entanglement can still be distributed. This is found in separable
Gaussian environments, where distillable entanglement is able to survive the
two-mode transmission, despite being broken in any single-mode transmission by
the strong thermal noise. In the Gaussian setting, entanglement restoration is
a threshold process, occurring only after a critical amount of correlations
has been injected. Such findings suggest new perspectives for distributing
entanglement in realistic environments with extreme decoherence, identifying
separable correlations and classical memory effects as physical resources for
\textquotedblleft breaking entanglement-breaking\textquotedblright.

\end{abstract}

\author{Stefano Pirandola}
\affiliation{Computer Science, University of York, York YO10 5GH, United Kingdom}
\maketitle


\section{Introduction}

Entanglement is a fundamental physical resource in quantum information and
computation~\cite{NielsenBook,Mwilde}. Once two remote parties, say Alice and
Bob, share a suitable amount of entanglement, they can implement a variety of
powerful protocols, including teleportation of quantum
states~\cite{tele1,tele1CV} and quantum gates~\cite{tele2}, and the
distribution of unconditionally secure keys~\cite{QKD,QKD2}. The problem of
entanglement distribution is therefore a central topic of investigation in the
quantum information community. Unfortunately, this distribution is also
fragile: Quantum systems inevitably interact with the external environment
whose decoherent action typically degrades their entanglement. In realistic
implementations where the effect of decoherence is non-negligible,
entanglement distribution may become challenging and may need distillation
protocols~\cite{Distillation,Distillation2}, where a large number of
weakly-entangled states are converted into fewer but more entangled states.

The worst case scenario is when decoherence is so strong as to destroy any
input entanglement. Mathematically, this is represented by the concept of
entanglement-breaking channel~\cite{EBchannels,HolevoEB}. In general, a
quantum channel $\mathcal{E}$\ is entanglement-breaking when its local action
on one part of a bipartite state always results into a separable output state,
no matter what the initial state were. In other words, given two systems, $A$
and $B$, in an arbitrary bipartite state $\rho_{AB}$, the output state
$\rho_{AB^{\prime}}=(\mathcal{I}_{A}\otimes\mathcal{E}_{B})(\rho_{AB})$\ is
always separable, where $\mathcal{I}_{A}$ is the identity channel applied to
system $A$ and $\mathcal{E}_{B}$ is the entanglement-breaking channel applied
to system $B$.

\begin{figure}[ptbh]
\vspace{-2.7cm}
\par
\begin{center}
\includegraphics[width=0.65\textwidth] {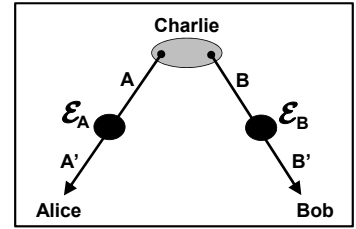}
\end{center}
\par
\vspace{-2.8cm}\caption{Symmetric distribution in a memoryless environment.
Charlie is the middle station with an entangled state of systems $A$ and $B$.
The communication lines with Alice and Bob are affected by two independent
entanglement-breaking channels, $\mathcal{E}_{A}$ and $\mathcal{E}_{B}$. In
this environment, no entanglement can be distributed, neither via single
transmission (Charlie $\rightarrow$ Alice or Bob), nor via double (Charlie
$\rightarrow$ Alice and Bob).}%
\label{scenario}%
\end{figure}

Despite entanglement-breaking channels having been the subject of an intensive
study by the community, they have only been analyzed under Markovian
conditions of no memory. In other words, when the distribution involves two or
more systems, these systems are typically assumed to be perturbed in an
independent fashion, each of them subject to the same memoryless channel. For
instance, consider the symmetric scheme in Fig.~\ref{scenario}, where a middle
station (Charlie) has a bipartite system $AB$ in some entangled state, but its
communication lines with two remote parties (Alice and Bob) are affected by
entanglement-breaking channels $\mathcal{E}_{A}$ and $\mathcal{E}_{B}$.

Under the assumption of memoryless channels, there is clearly no way to
distribute entanglement among any of the parties. Suppose that Charlie tries
to share entanglement with one of the remote parties by sending one of the two
systems while keeping the other (a scenario that we call \textquotedblleft%
1-system transmission\textquotedblright\ or just \textquotedblleft single
transmission\textquotedblright). For instance, Charlie may keep system $A$
while transmitting system $B$ to Bob. The action of $\mathcal{I}_{A}%
\otimes\mathcal{E}_{B}$ destroys the initial entanglement, so that systems $A$
(kept) and $B^{\prime}$ (transmitted) are separable. Symmetrically, the action
of $\mathcal{E}_{A}\otimes\mathcal{I}_{B}$ destroys the entanglement between
system $A^{\prime}$ (transmitted) and system $B$ (kept). Now suppose that
Charlie sends system $A$ to Alice and system $B$ to Bob (a scenario that we
call \textquotedblleft2-system transmission\textquotedblright\ or
\textquotedblleft double transmission\textquotedblright). Since the joint
action of the two\ channels is given by the tensor product $\mathcal{E}%
_{A}\otimes\mathcal{E}_{B}=(\mathcal{E}_{A}\otimes\mathcal{I}_{B}%
)(\mathcal{I}_{A}\otimes\mathcal{E}_{B})$ quantum entanglement must
necessarily be destroyed. In other words, since we have 1-system
entanglement-breaking, then we must have 2-system entanglement-breaking.

In this paper we discuss how the previous implication is false when we
introduce correlations, i.e., a memory, between the two entanglement-breaking
channels. In the presence of a correlated-noise environment, the double
transmission can successfully distribute entanglement despite the single
transmission being subject to entanglement-breaking: Charlie can transmit
entanglement to Alice and Bob, despite not being able to share any
entanglement with them. Surprisingly, this effect can be induced by the
presence of separable correlations in the joint environment, so that the
broken entanglement is restored in a subtle way and it is not just replaced by
other entanglement coming from the environment. In other words, to achieve
this effect we do not need to consider arbitrary joint dilations of the two
channels, but just two independent unitary dilations which are coupled by a
separable environmental state.\begin{figure}[ptbh]
\vspace{-2.4cm}
\par
\begin{center}
\includegraphics[width=0.62\textwidth] {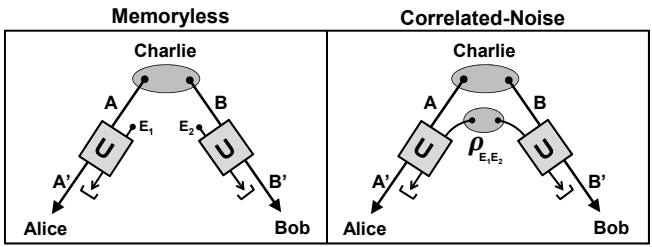}
\end{center}
\par
\vspace{-2.7cm}\caption{\textit{Left panel}.~Symmetric distribution in a
memoryless environment. The two entanglement-breaking channels, $\mathcal{E}%
_{A}$ and $\mathcal{E}_{B}$, are dilated in two independent unitaries,
$U_{E_{1}A}$ and $U_{E_{2}B}$, combining the input systems with two
environmental systems, $E_{1}$ and $E_{2}$, described by a product-state
$\rho_{E_{1}}\otimes\rho_{E_{2}}$ (see text for more details). \textit{Right
panel}.~Symmetric distribution in a correlated-noise environment. Here we
create correlations between the two entanglement-breaking channels. The
cheapest way is to replace $\rho_{E_{1}}\otimes\rho_{E_{2}}$ with a separable
state $\rho_{E_{1}E_{2}}$. In this scenario, despite entanglement cannot be
distributed by single transmissions (Charlie cannot be entangled with Alice or
Bob), still it can be distributed via the double transmission (so that Alice
and Bob can become entangled). This reactivation of entanglement distribution
is induced by the separable correlations injected from the environment. For
twirling environments (in any dimension), the separable state $\rho
_{E_{1}E_{2}}$ is also classical (zero-discord), so that entanglement
reactivation is mediated by purely-classical correlations.}%
\label{scenario2}%
\end{figure}

To better clarify these points, consider Fig.~\ref{scenario2}. In the left
panel, the two memoryless entanglement-breaking channels, $\mathcal{E}_{A}$
and $\mathcal{E}_{B}$, are dilated in two unitaries, $U_{E_{1}A}$ and
$U_{E_{2}B}$, coupling the input systems, $A$ and $B$, with environmental
systems, $E_{1}$ and $E_{2}$, described by a product-state $\rho_{E_{1}%
}\otimes\rho_{E_{2}}$. In other words, the reduced state $\rho_{A}%
=\mathrm{Tr}_{B}(\rho_{AB})$ is transformed by the map%
\begin{equation}
\mathcal{E}_{A}(\rho_{A})=\mathrm{Tr}_{E_{1}}[U_{E_{1}A}(\rho_{E_{1}}%
\otimes\rho_{A})U_{E_{1}A}^{\dagger}],
\end{equation}
and, similarly, $\rho_{B}=\mathrm{Tr}_{A}(\rho_{AB})$ is transformed as%
\begin{equation}
\mathcal{E}_{B}(\rho_{B})=\mathrm{Tr}_{E_{2}}[U_{E_{2}B}(\rho_{E_{2}}%
\otimes\rho_{B})U_{E_{2}B}^{\dagger}].
\end{equation}

Now, we note that there are many ways in which we can correlate the two
channels. One could consider a completely general joint unitary
$U_{AB\mathbf{E}}$ involving both the input systems, together with ancillary
systems $\mathbf{E}=E_{1}E_{2}\ldots$ However this general approach is not
interesting and somehow trivial, since it always includes cases where the
input entanglement is broken ($\rho_{A^{\prime}B}$ and $\rho_{AB^{\prime}}$
both separable) and replaced by fresh entanglement coming from the environment
(so that $\rho_{A^{\prime}B^{\prime}}$ is entangled).

More interestingly, our analysis regards more realistic scenarios where the
environment has minimal resources, i.e., it is weakly correlated, therefore
far from being entangled. This is interesting for its potential applications
to realistic non-Markovian systems, and theoretically non-trivial, since no
swapping of entanglement can occur between systems and environment. In other
words, we are interested in studying the weakest models of correlated-noise
environment, close to the memoryless paradigm, which are able to reactivate
the entanglement distribution.

Driven by such a goal, the simplest and cheapest way to combine two
entanglement breaking channels and create a weakly-correlated joint
environment, is to keep the previous independent unitaries, $U_{E_{1}A}$ and
$U_{E_{2}B}$, and replace the environmental product-state $\rho_{E_{1}}%
\otimes\rho_{E_{2}}$ with a separable state $\rho_{E_{1}E_{2}}$ having the
same marginals $\rho_{E_{1}}=\mathrm{Tr}_{E_{2}}(\rho_{E_{1}E_{2}})$ and
$\rho_{E_{2}}=\mathrm{Tr}_{E_{1}}(\rho_{E_{1}E_{2}})$, but correlated
$\rho_{E_{1}E_{2}}\neq\rho_{E_{1}}\otimes\rho_{E_{2}}$ (see the right panel of
Fig.~\ref{scenario2}). Assuming this minimal dilation, we create a joint
channel $\mathcal{E}=\mathcal{E}_{AB}$, transforming the input state
$\rho_{AB}$ as%
\begin{equation}
\mathcal{E}(\rho_{AB})=\mathrm{Tr}_{E_{1}E_{2}}[W(\rho_{E_{1}E_{2}}\otimes
\rho_{AB})W^{\dagger}]~,\label{eqDIL}%
\end{equation}
where $W:=U_{E_{1}A}\otimes U_{E_{2}B}$. Clearly, the joint channel is not
memoryless $\mathcal{E}\neq\mathcal{E}_{A}\otimes\mathcal{E}_{B}$. Also note
that the two local channels, $\mathcal{E}_{A}$ and $\mathcal{E}_{B}$, are
still well defined. For instance, if only system $A$ is transmitted, we have%
\begin{align}
\rho_{A^{\prime}B}  & =(\mathcal{E}_{A}\otimes\mathcal{I}_{B})(\rho_{AB})\\
& =\mathrm{Tr}_{E_{1}}[(U_{E_{1}A}\otimes I_{B})(\rho_{E_{1}}\otimes\rho
_{AB})(U_{E_{1}A}^{\dagger}\otimes I_{B})].\nonumber
\end{align}

In our work we show that, despite $\mathcal{E}_{A}$ and $\mathcal{E}_{B}$
being entanglement-breaking ($\rho_{A^{\prime}B}$ and $\rho_{AB^{\prime}}$
separable for any input $\rho_{AB}$), the joint channel $\mathcal{E}$ is able
to transmit entanglement, in the sense that $\rho_{A^{\prime}B^{\prime}}$ can
be entangled for a suitable choice of the input entangled state $\rho_{AB}$.
This entanglement reactivation is clearly an effect of the injected
correlations, coming from the separable state of the environment $\rho
_{E_{1}E_{2}}$. As mentioned earlier, the interesting point is thar the
distribution is restored as an effect of weak, local-type correlations.

The reactivation of entanglement occurs in different kinds of separable
environments, with rather different physical properties. The simplest
environments to construct are the \textquotedblleft twirling
environments\textquotedblright, which are based on the twirling operators
$U\otimes U$ (or $U\otimes U^{\ast}$). In this case, a random unitary $U$ is
directly applied to system $A$ and the same unitary (or its conjugate) is
correspondingly applied to system $B$. While the local action of a random
unitary ($U\otimes I$ or $I\otimes U$) is entanglement-breaking, the
correlated action ($U\otimes U$) perfectly preserves specific classes of
entangled states, belonging to an invariant decoherent-free subspace of the
joint correlated environment~\cite{NOTAdeco}. Since these environments\ are
based on local operations and classical communication (LOCC), they are
expected to introduce correlations which are separable and, more precisely,
purely-classical. This feature can be explicitly checked by performing the
unitary dilation of these environments according to Eq.~(\ref{eqDIL}), and
checking that the environmental state $\rho_{E_{1}E_{2}}$ is a classical
state, therefore, with zero quantum discord.

Twirling environments can easily be constructed for quantum systems with
Hilbert spaces of any dimension, both finite (qudits) and infinite
(continuous-variable systems~\cite{BraREV2,RMP}). In the case of two qudits,
the randomization of the twirling operators $U\otimes U$ (or $U\otimes
U^{\ast}$) is performed over the entire unitary group. In this case, it is
easy to identify states which are invariant under $U\otimes U$-twirling
(Werner states~\cite{Werner}) and $U\otimes U^{\ast}$-twirling (isotropic
states~\cite{HOROs}). In the specific case of qubits ($d=2$), we can restrict
the randomization to the basis of the Pauli operators, with the qubit Werner
state being invariant under Pauli twirling.

Things are less trivial for infinite dimension, in particular, for bosonic
modes~\cite{BraREV2,RMP}. In this case, we restrict the twirling operators to
the compact group of orthogonal symplectic transformations, i.e., phase-space
rotations. The bosonic twirling environment so defined is non-Gaussian. Here,
it is interesting to see that, only for the $U\otimes U^{\ast}$-twirling, we
can identify Gaussian states which are invariant and entangled: These are the
two mode squeezed vacuum states, also known as Einstein-Podolsky-Rosen (EPR)
states~\cite{EPR,RMP}. More generally, this class can be extended to
continuous variable Werner states, which are non-Gaussian states given by
mixing an EPR state with a tensor product of thermal states~\cite{CVwerner}.

Thanks to the existence of an invariant subspace of entangled states, twirling
channels allow for the perfect distribution of entanglement in the two-system
transmission, despite being one-system entanglement-breaking. However, the
presence of such a subspace is not necessary if we relax the condition of
perfect transmission and we accept that only a fraction of the input
entanglement survives the two-system transmission. Remarkably, this is
possible for bosonic systems evolving in a realistic model of correlated-noise
Gaussian environment, which is the direct generalization of the standard
thermal-lossy environment.

Our model of Gaussian environment can be represented by two beam-splitters
($U_{E_{1}A}$ and $U_{E_{2}B}$) whose environmental ports are subject to a
correlated (separable) two-mode thermal state $\rho_{E_{1}E_{2}}$. Such an
environment does not have decoherence-free subspaces, apart from the trivial
invariant space given by its own thermal state $\rho_{E_{1}E_{2}}$. Then, for
a sufficiently large amount of entanglement at the input, Charlie can
distribute a distillable amount to Alice and Bob, despite the single quantum
channels Charlie-Alice ($\mathcal{E}_{A}$) and Charlie-Bob ($\mathcal{E}_{B}$)
being entanglement-breaking. In particular, we find a threshold behavior
according to which remote entanglement is restored only after a critical
amount of correlations has been injected by the environment. Once this
reactivation has taken place, we may ask if the amount of remote entanglement
is then increasing in the amount of injected correlations. Unfortunately such
a simple monotonicity does not seem to hold in the general case.

In fact, for these Gaussian environments, we adopt Gaussian discord as a
measure of quantum correlations (upper bound to the actual discord). Then, we
can easily evaluate the number of entanglement bits (ebits) which are remotely
distributed in terms of the number of discordant bits (dbits)\ and classical
bits (cbits), which are injected by the environment. For some classes of
Gaussian environments, remote entanglement is proven to be increasing in the
injected correlations (both classical and quantum). However, this monotonic
behavior is not a general property, since we can exhibit two examples of
environments, with identical classical correlations and increasing Gaussian
discord, for which the number of ebits remotely restored is strictly
decreasing. Because of this non-monotonic behavior, the phenomenon of
entanglement restoration in Gaussian environments does not seem to be easily
or directly related to a specific quantification of the correlations
(classical, quantum or total).


To clarify relations with previous literature, we remark that our study is
specifically devoted to show the potentialities of separable correlations in
breaking the mechanism of entanglement-breaking, so that an entangled state is
able to keep some of the initial entanglement during the transmission towards
remote parties. It is clear that this is different from what considered in
previous studies~\cite{old1,old2,old3,old4}, where separable probes were used
to distribute \textquotedblleft fresh entanglement\textquotedblright\ between
two quantum systems that have never interacted before. This basic difference
leads to completely different roles for quantum discord. While in
Refs.~\cite{old1,old2,old3,old4} discord is identified as the fundamental
physical resource needed to distribute fresh entanglement, in our study
discord is not necessary in general, since purely-classical correlations can
indeed be sufficient to restore broken entanglement (e.g., see twirling environments).

In terms of potential impact, our work opens new possibilities for
entanglement distribution in environments with extreme decoherence, where the
presence of noise correlations and memory effects can be exploited to recover
from entanglement-breaking. Since entanglement restoration can be achieved by
separable correlations, our work also poses fundamental questions on the
intimate relations between local and nonlocal correlations and, more
generally, between classical and quantum correlations. It is important to note
that memory channels and correlated (in particular, non-Markovian)
environments are present in a wide series of practical
scenarios~\cite{CosmoREV}, including spin chains~\cite{Bose03}, atoms in
optical cavities~\cite{Benenti09}, quantum dots in photonic
crystals~\cite{Madsen11}, photonic propagation through linear optical
systems~\cite{Shapiro09,Lupo1,Lupo2,qreadDIFF} and atmospheric
turbulence~\cite{Tyler09,Semenov09,Boyd11}.

The general structure of the paper is the following. In Sec.~\ref{SECtwirl},
we study twirling environments showing how entanglement is perfectly restored
by classical correlations. This is proven for quantum systems of any dimension
(qubits, qudits and bosonic systems). In Sec.~\ref{SECgaussMAIN}, we consider
bosonic Gaussian environments. Here distillable entanglement can be restored
by separable correlations, whose composition in terms of classical
correlations and Gaussian discord is also analyzed. Finally,
Sec.~\ref{SECconclusion} is for conclusion and discussion. For the sake of
completeness, we also include an Appendix~\ref{APPmisce}, containing simple
technical proofs, and an Appendix~\ref{APPbosonic}, which is a brief review on
bosonic systems and Gaussian states.

\section{Entanglement preservation in twirling environments\label{SECtwirl}}

In this first analysis, we combine two entanglement-breaking channels to form
a twirling environment, where entanglement distribution is reactivated by the
presence of classical correlations. In particular, two-system transmission
from Charlie to Alice and Bob allows for a perfect transfer of entanglement,
thanks to the existence of an invariant subspace of entangled states. As
already said, this is proven for Hilbert spaces of any dimension. In
Sec.~\ref{QubitSEC}, we start by considering the transmission of qubits in the
simplest examples of twirling environments (correlated Pauli environments).
Then, we consider the general case of qudits evolving in multidimensional
twirling environments (Sec.~\ref{TwirlSECTION}). Finally, we consider bosonic
systems subject to non-Gaussian twirling environments which are based on
random phase-space rotations (Sec.~\ref{BOSsections}).

\subsection{Qubits in correlated Pauli environments\label{QubitSEC}}

The simplest example can be constructed for qubits considering the basis of
the four Pauli operators $I$, $X$, $Y=iXZ$ and $Z$~\cite{NielsenBook}. Given
an arbitrary two-qubit state $\rho_{AB}$, we consider the correlated Pauli
channel
\begin{equation}
\mathcal{E}(\rho_{AB})=\sum_{k=0}^{3}p_{k}~(P_{k}\otimes P_{k})\rho_{AB}%
(P_{k}\otimes P_{k})^{\dagger}~, \label{qubitENV}%
\end{equation}
where $P_{k}\in\{I,X,Y,Z\}$ and $p_{k}$ are probabilities. This channel is
clearly simulated by random LOCCs. In fact, it is equivalent to extract a
random variable $K=\{k,p_{k}\}$, apply the Pauli unitary $P_{k}$\ to qubit $A
$, communicate $k$ and then apply the same unitary $P_{k}$\ to qubit $B$. It
is easy to check that the two-qubit channel of Eq.~(\ref{qubitENV}) does not
change if we replace the Pauli twirling operator $P_{k}\otimes P_{k}$ with the
alternative operator $P_{k}\otimes P_{k}^{\ast}$.

It is easy to write the unitary dilation of the correlated Pauli channel. It
is sufficient to introduce an environment composed by two systems $E_{1}$ and
$E_{2}$, each being a qudit with dimension $d=4$ and orthonormal basis
$\{\left\vert k\right\rangle \}_{k=0}^{3}$. Then, we can write the dilation of
Eq.~(\ref{eqDIL}), where the environment is prepared in the correlated state%
\begin{equation}
\rho_{E_{1}E_{2}}=\sum_{k=0}^{3}p_{k}\left\vert k\right\rangle _{E_{1}%
}\left\langle k\right\vert \otimes\left\vert k\right\rangle _{E_{2}%
}\left\langle k\right\vert ~, \label{StateENV1}%
\end{equation}
and the unitary interactions are control-Pauli unitaries,
\begin{equation}
U_{E_{1}A}=\sum_{k=0}^{3}\left\vert k\right\rangle _{E_{1}}\left\langle
k\right\vert \otimes P_{k},~U_{E_{2}B}=\sum_{k=0}^{3}\left\vert k\right\rangle
_{E_{2}}\left\langle k\right\vert \otimes P_{k}. \label{Cunitaries}%
\end{equation}
As evident from Eq.~(\ref{StateENV1}), the state of the environment is
separable, which means that only separable (i.e., local-type) correlations are
injected into the travelling qubits. More precisely, since $\rho_{E_{1}E_{2}}$
is expressed as a convex combination of orthogonal projectors, it is a
purely-classical state, i.e., a state with zero quantum discord~\cite{RMPdis}.
As a result, this environment contains correlations which are not only local
but also purely classical.

We now show the conditions under which the correlated Pauli environment is
simultaneously one-qubit entanglement-breaking and two-qubit
entanglement-preserving. We start by considering the transmission of one qubit
only, e.g., qubit $A$. This is subject to a depolarizing channel
\begin{equation}
(\mathcal{E}_{A}\otimes\mathcal{I}_{B})(\rho_{AB})=p_{0}\rho_{AB}+\sum
_{k=1}^{3}p_{k}(P_{k}\otimes I)\rho_{AB}(P_{k}^{\dagger}\otimes I).
\label{qubitDEPO}%
\end{equation}
It is easy to show that $\mathcal{E}_{A}$\ is entanglement-breaking when
$p_{k}\leq1/2$ for every $k$ (see Appendix~\ref{DEPapp}\ for a simple proof).
A particular choice can be $p_{0}\leq1/2$ and $p_{1}=p_{2}=p_{3}=(1-p_{0})/3$
as for instance used in Ref.~\cite{Sacchi}.

Assuming the condition of one-qubit entanglement-breaking\ ($p_{k}\leq1/2$),
Charlie is clearly not able to share any entanglement with Alice or Bob.
Despite this, Charlie can still distribute entanglement to Alice and Bob. This
is possible because we can identify a class of entangled states $\rho_{AB}$
which are invariant under the action of the correlated map~(\ref{qubitENV}),
i.e., $\mathcal{E}(\rho_{AB})=\rho_{AB}$. This class is simply given by the
Werner states%
\begin{equation}
\rho_{AB}(\gamma):=(1-\gamma)\frac{I_{AB}}{4}+\gamma\left\vert -\right\rangle
_{AB}\left\langle -\right\vert ~, \label{2qubitWERNER}%
\end{equation}
with parameter $-1/3\leq\gamma\leq1$, where $I_{AB}/4$ is the maximally-mixed
state and
\begin{equation}
\left\vert -\right\rangle _{AB}=\frac{1}{\sqrt{2}}\left(  \left\vert
0\right\rangle _{A}\left\vert 1\right\rangle _{B}-\left\vert 1\right\rangle
_{A}\left\vert 0\right\rangle _{B}\right)
\end{equation}
is the maximally-entangled (singlet) state. For $\gamma>1/3$, the state
$\rho_{AB}(\gamma)$ is known to be entangled and distillable. The remarkable
property of the Werner states is that they are invariant under twirling
operators $U\otimes U$, i.e.,%
\begin{equation}
(U\otimes U)\rho_{AB}(\gamma)(U\otimes U)^{\dagger}=\rho_{AB}(\gamma)~,
\end{equation}
for any unitary $U$. As a result, they are clearly invariant under the action
of the correlated Pauli environment of Eq.~(\ref{qubitENV}). Thus, if Charlie
sends Werner states with $\gamma>1/3$, these entangled states are perfectly
distributed to Alice and Bob (two-qubit entanglement-preserving).

\subsection{Qudits in multidimensional twirling
environments\label{TwirlSECTION}}

In this section, we consider the general case of qudits, i.e., quantum systems
with Hilbert space of arbitrary dimension $d\geq2$. For these systems, we can
easily construct classically-correlated environments which are simultaneously
1-qudit entanglement-breaking and 2-qudit entanglement-preserving.

Consider two qudits, $A$ and $B$, with Hilbert spaces of the same dimension
$d$ and prepared in a bipartite state $\rho_{AB}$. A multidimensional
$U\otimes U$ twirling channel is described by the following
completely-positive trace-preserving map%
\begin{equation}
\mathcal{E}_{UU}(\rho_{AB})=\int_{\mathcal{U}(d)}dU~(U\otimes U)\rho
_{AB}(U\otimes U)^{\dagger}~, \label{GENtwMAP}%
\end{equation}
where the integral is over the entire unitary group $\mathcal{U}(d)$ acting on
the $d$-dimensional Hilbert space, and $dU$ is the Haar measure. This channel
is clearly realizable by random LOCCs: A random unitary $U$ is drawn and
applied to qudit $A$, the choice of $U$ is classically communicated to the
other qudit $B$, which is then subject to the same unitary. Similarly, we can
define a $U\otimes U^{\ast}$ twirling channel, by replacing the twirling
operator $U\otimes U$ with the alternative twirling $U\otimes U^{\ast}$ in the
definition of Eq.~(\ref{GENtwMAP}). Compactly, we refer to the $U\otimes V$
twirling channel%
\begin{equation}
\mathcal{E}_{UV}(\rho_{AB})=\int_{\mathcal{U}(d)}dU~(U\otimes V)\rho
_{AB}(U\otimes V)^{\dagger}~, \label{genTWIRL}%
\end{equation}
where $V=U$ or $V=U^{\ast}$.

This channel can be dilated to introduce an environmental system. Since we
have an integral in Eq.~(\ref{genTWIRL}), this dilation seems to involve the
introduction of continuous variable systems. In fact, the unitary group
$\mathcal{U}(d)$ is labelled by $d^{2}$ real parameters~\cite{UnitaryPAR},
which means that $2d^{2}$ continuous variable systems are needed to embed
these parameters and describe the environment. Actually, such a continuous
dilation is not necessary, since we can always replace the previous Haar
integral with a discrete sum over a finite number of suitably-chosen
unitaries. In fact, any twirling channel~(\ref{genTWIRL}) can be written as%
\begin{equation}
\mathcal{E}_{UV}(\rho_{AB})=\frac{1}{K}\sum_{k=0}^{K-1}(U_{k}\otimes
V_{k})\rho_{AB}(U_{k}\otimes V_{k})^{\dagger}~, \label{design1}%
\end{equation}
where $U_{k}$ belongs to the set of unitary 2-design $\mathcal{D}%
$~\cite{Designs,Des2} and $V_{k}=U_{k}$ or $V_{k}=U_{k}^{\ast}$. The set
$\mathcal{D}$ has a finite number of elements which depends of the dimension
of the Hilbert space $K=K(d)$ (to see how the cardinality $K$ scales with the
dimension $d$, see Ref.~\cite{Gross}). The proof of the equivalence between
Eqs.~(\ref{genTWIRL}) and~(\ref{design1}) can be found in Ref.~\cite{Designs}
for the $U\otimes U$ twirling channel. See Appendix~\ref{TwirlAVEapp}\ for a
simple extension of the proof to the other $U\otimes U^{\ast}$ twirling
channel. Note that, in the case of qubits ($d=2$), an example of unitary
2-design is provided by the Clifford group~\cite{Designs,Clifford3}, which is
the normalizer of the Pauli group and typically employed in quantum error
correction~\cite{Clifford1,Clifford2}.

Now using the unitary 2-design, we can dilate the twirling channel into an
environment made by finite-dimensional systems, i.e., two larger qudits
$E_{1}$ and $E_{2}$, each with dimension $K$ and orthonormal basis
$\{\left\vert k\right\rangle \}_{k=0}^{K-1}$. In other words, for
$\mathcal{E}_{UV}(\rho_{AB})$ we can write the dilation of Eq.~(\ref{eqDIL}),
where the environment is prepared in the uniformly-correlated state%
\begin{equation}
\rho_{E_{1}E_{2}}=\frac{1}{K}\sum_{k=0}^{K-1}\left\vert k\right\rangle
_{E_{1}}\left\langle k\right\vert \otimes\left\vert k\right\rangle _{E_{2}%
}\left\langle k\right\vert ~, \label{StateENV2}%
\end{equation}
and the interactions are the following control-unitaries
\begin{equation}
U_{E_{1}A}=\sum_{k=0}^{K-1}\left\vert k\right\rangle _{E_{1}}\left\langle
k\right\vert \otimes U_{k},~U_{E_{2}B}=\sum_{k=0}^{K-1}\left\vert
k\right\rangle _{E_{2}}\left\langle k\right\vert \otimes V_{k},
\label{Cunitaries2}%
\end{equation}
where $U_{k}\in\mathcal{D}$ and $V_{k}=U_{k}$ or $V_{k}=U_{k}^{\ast}$. As
evident from Eq.~(\ref{StateENV2}), the state of the environment is separable
and purely-classical (zero discord). As expected, twirling environments only
contain purely-classical correlations.

Once we have fully characterized the properties of these environments, we show
the conditions under which they are simultaneously one-qudit
entanglement-breaking and two-qudit entanglement-preserving. First of all, we
explicitly show that one-qudit transmission is always subject to
entanglement-breaking. In fact, suppose that Charlie transmits qudit $A$ to
Alice while keeping qudit $B$. For any input state $\rho_{AB}$, the output
state is given by%
\begin{align}
(\mathcal{E}_{A}\otimes\mathcal{I}_{B})(\rho_{AB})  &  =\nonumber\\
\int_{\mathcal{U}(d)}dU~(U\otimes I)\rho_{AB}(U^{\dagger}\otimes I)  &
=\frac{I}{d}\otimes\mathrm{Tr}_{A}(\rho_{AB})~. \label{depola2}%
\end{align}
It is clear that the random map $U\otimes I$ implements an
entanglement-breaking channel (a symmetric result holds for the other random
map $I\otimes V$ which describes the transmission of qudit $B$). As shown in
Appendix~\ref{HaarAVEsec}, the proof of Eq.~(\ref{depola2}) is an application
of the identity
\begin{equation}
\left\langle O\right\rangle _{U}:=\int_{\mathcal{U}(d)}dU~UOU^{\dagger}%
=\frac{\mathrm{Tr}(O)}{d}I~, \label{twirlingID}%
\end{equation}
which is the Haar average of a linear operator $O$. In turn,
Eq.~(\ref{twirlingID}) is a consequence of Schur's lemma and the invariance of
the Haar measure~\cite{Chiri}.

The next step is the analysis of the two-qudit transmission, from Charlie to
Alice and Bob. In this case, we look for entangled states which are preserved
by the correlated action of the twirling environment. Luckily, we can easily
find states which are invariant under the action of the twirling operator
$U\otimes V$, i.e.,
\begin{equation}
(U\otimes V)\rho_{AB}(U\otimes V)^{\dagger}=\rho_{AB}~. \label{TwirlINVA}%
\end{equation}
Thanks to this invariance, such states are fixed points of the $U\otimes V$
twirling channel of Eq.~(\ref{genTWIRL}).

In the specific case of $V=U$, it is well known that the unique solution of
Eq.~(\ref{TwirlINVA}) is provided by the multidimensional Werner
states~\cite{Werner}. For two qudits $A$ and $B$ (with the same dimension
$d$), the Werner states are defined by the one-parameter
class~\cite{Werner,Synak}%
\begin{equation}
\rho_{AB}(\mu):=\frac{1}{d^{2}+d\mu}(I_{AB}+\mu V) \label{multiWerner}%
\end{equation}
where $-1\leq\mu\leq1$, and $V$ is the unitary flip operator $V\left\vert
\varphi\right\rangle _{A}\otimes\left\vert \psi\right\rangle _{B}=\left\vert
\psi\right\rangle _{A}\otimes\left\vert \varphi\right\rangle _{B}$. These
states are known to be entangled and distillable for any physical value of
$\mu<-d^{-1}$. Thus, if Charlie has a Werner state $\rho_{AB}(\mu)$ with
suitable $\mu$ (in the entanglement regime), he is able to perfectly transmit
this state to Alice and Bob, who can therefore share and distill entanglement.
Such a distribution is possible, despite Charlie not being able to share any
entanglement with Alice or Bob due to the entanglement-breaking condition of
Eq.~(\ref{depola2}).

Coming back to Eq. (\ref{TwirlINVA}), one can find a similar solution for
$V=U^{\ast}$. In fact, as shown in Ref.~\cite{HOROs}, there exist states which
are invariant under $U\otimes U^{\ast}$-twirling. These are called isotropic
states, and they are defined by the one-parameter class~\cite{HOROs}%
\begin{equation}
\rho_{AB}(\gamma):=(1-\gamma)\frac{I_{AB}}{d^{2}}+\gamma\left\vert
\psi\right\rangle _{AB}\left\langle \psi\right\vert ~, \label{Isostates}%
\end{equation}
involving a convex combination of the maximally mixed state $d^{-2}I_{AB}%
$\ and the maximally entangled state
\begin{equation}
\left\vert \psi\right\rangle _{AB}=\frac{1}{\sqrt{d}}\sum_{k=0}^{d-1}%
\left\vert k\right\rangle _{A}\left\vert k\right\rangle _{B}~,
\label{maxENTstate}%
\end{equation}
with parameter $-(d^{2}-1)^{-1}\leq\gamma\leq1$. In general, these states are
entangled and distillable for any physical value of $\gamma>(1+d)^{-1}$. Thus,
if Charlie has an isotropic state $\rho_{AB}(\gamma)$ with suitable $\gamma$
(i.e., in the entanglement regime), he is able to perfectly transmit this
state to Alice and Bob, who therefore can share and distill entanglement.
Again, this is possible despite the transmission of a single qudit being
subject to an entanglement-breaking channel according to Eq.~(\ref{depola2}).

It is clear that these results can be specialized to the case of qubits ($d=2
$). For qubits, the classes of multidimensional Werner states of
Eq.~(\ref{multiWerner}) and isotropic states of Eq.~(\ref{Isostates}) coincide
up to a local unitary~\cite{HOROs}. Multidimensional Werner states reduce
exactly to the qubit Werner state of Eq.~(\ref{2qubitWERNER}) which is
$U\otimes U$-invariant~\footnote{In fact, for $d=2$, we have $(I_{AB}+\mu
V)=(\mu+1)I_{AB}-2\mu\left\vert -\right\rangle _{AB}\left\langle -\right\vert
$ which, replaced in Eq.~(\ref{multiWerner}), gives the state of
Eq.~(\ref{2qubitWERNER}) by setting $\gamma=-\mu/(2+\mu)$.}. On the other
hand, isotropic states reduce to Eq.~(\ref{2qubitWERNER}), proviso that the
singlet $\left\vert -\right\rangle _{AB}$ is replaced by the triplet
\begin{equation}
\left\vert +\right\rangle _{AB}=\frac{1}{\sqrt{2}}\left(  \left\vert
0\right\rangle _{A}\left\vert 0\right\rangle _{B}+\left\vert 1\right\rangle
_{A}\left\vert 1\right\rangle _{B}\right)  ~. \label{tripletSTATE}%
\end{equation}
This state is now $U\otimes U^{\ast}$-invariant and known as Werner-like state
(this definition is commonly adopted when the singlet is replaced by another
maximally entangled state). If we restrict the random unitaries to the basis
of the Pauli operators $\{P_{k}\}$ as done in Sec.~\ref{QubitSEC}, we see that
both these states are fixed points of the correlated Pauli environment of
Eq.~(\ref{qubitENV}), since this map does not change under the replacement
$P_{k}\otimes P_{k}\rightarrow P_{k}\otimes P_{k}^{\ast}$.

\subsection{Bosonic systems in non-Gaussian twirling
environments\label{BOSsections}}

Here we extend the analysis to the case of continuous variable systems, i.e.,
quantum systems with infinite dimensional Hilbert spaces ($d=\infty$). In
particular, we consider the case of two bosonic modes of the electromagnetic
field (see Appendix~\ref{APPbosonic}\ for a brief review of the main concepts
on bosonic systems). The simplest generalization of the notion of twirling
environment involves the use of rotations in the phase space. Given a single
bosonic mode with number operator $\hat{n}$, the rotation operator is defined
as $R_{\theta}=\exp(-i\theta\hat{n})$. In the phase space, the symplectic
action of this operator is described by the rotation matrix
\begin{equation}
\mathbf{R}_{\theta}=\left(
\begin{array}
[c]{cc}%
\cos\theta & \sin\theta\\
-\sin\theta & \cos\theta
\end{array}
\right)  ~.
\end{equation}
In terms of the second-order statistical moments, we have that the covariance
matrix (CM) $\mathbf{V}$ of the input mode is transformed by the congruence
\begin{equation}
\mathbf{V}\rightarrow\mathbf{R}_{\theta}\mathbf{VR}_{\theta}{}^{T}~.
\end{equation}

Now, given an input state $\rho_{AB}$ of two modes, $A$ and $B$, we can
synchronize two random rotations and define the bosonic twirling channel as%
\begin{equation}
\mathcal{E}_{\theta\theta^{\prime}}(\rho_{AB})=\int\frac{d\theta}{2\pi
}~(R_{\theta}\otimes R_{\theta^{\prime}})\rho_{AB}(R_{\theta}\otimes
R_{\theta^{\prime}})^{\dagger}, \label{ROTchannel}%
\end{equation}
where $R_{\theta^{\prime}}=R_{\theta}$ or $R_{\theta^{\prime}}=R_{-\theta
}=R_{\theta}^{\ast}$. This is a non-Gaussian channel, since the twirling
operator $R_{\theta}\otimes R_{\theta^{\prime}}$, despite Gaussian, is
averaged using a uniform (i.e., non-Gaussian) distribution. It is clearly
based on random LOCCs, since random rotations are locally applied to each
bosonic mode and can be correlated by means of classical communication.

It is interesting that the unitary dilation of this channel can be restricted
to a finite-dimensional environment. This is because the single-mode rotation
operator $R_{\theta}$ belongs to the compact subgroup of the orthogonal
symplectic transformations $\mathcal{K}(2)=\mathcal{S}p(2)\cap\mathcal{O}(2)$
(these correspond to passive Gaussian unitaries, i.e., unitary transformations
preserving both the Gaussian statistics and the energy of the state). Then,
$\mathcal{K}(2)$ is isomorphic to the unitary group $\mathcal{U}(1)$, which is
the multiplicative group composed by all complex numbers with unit module,
also known as the \textquotedblleft circle group\textquotedblright.

Because of this isomorphism, a unitary 2-design $\mathcal{D}\subset
\mathcal{U}(1)$ can be mapped into a unitary 2-design for $\mathcal{K}%
(2)$~\cite{Gross}. As a result, we can write
\begin{equation}
\mathcal{E}_{\theta\theta^{\prime}}(\rho_{AB})=\frac{1}{K}\sum_{k=0}%
^{K-1}(R_{\theta_{k}}\otimes R_{\theta_{k}^{\prime}})\rho_{AB}(R_{\theta_{k}%
}\otimes R_{\theta_{k}^{\prime}})^{\dagger},
\end{equation}
for a suitable set of angles $\{\theta_{0},\ldots,\theta_{K-1}\}$, and
$\theta_{k}^{\prime}=\theta_{k}$ or $\theta_{k}^{\prime}=-\theta_{k}$. In this
form, the channel is manifestly non-Gaussian. In particular, it can be
represented by introducing two environmental qudits, $E_{1}$ and $E_{2}$,
prepared in a correlated state $\rho_{E_{1}E_{2}}$ as in Eq.~(\ref{StateENV2}%
), and interacting with the two bosonic modes by two control-unitaries as in
Eq.~(\ref{Cunitaries2}), where now the target unitaries, $U_{k}$ and $V_{k}$,
are rotations in the phase space. The environmental state is not only
separable but also purely-classical (zero discord), which means that only
classical correlations are injected by the bosonic twirling environment.

It is easy to show that one-mode transmission is always subject to
entanglement-breaking in this kind of environment. For instance, if mode $A$
is transmitted from Charlie to Alice, then the output state
\begin{equation}
\rho_{A^{\prime}B}=\int\frac{d\theta}{2\pi}~(R_{\theta}\otimes I)\rho
_{AB}(R_{\theta}\otimes I)^{\dagger} \label{UNIdepha}%
\end{equation}
is separable, no matter what the input state $\rho_{AB}$ is. In fact, it is
easy to prove that a uniformly dephasing channel as in Eq.~(\ref{UNIdepha}) is
entanglement-breaking. For completeness, we give this simple proof in
Appendix~\ref{DephaseAPP}.

The next step is to find two-mode states which are invariant under bosonic
twirling%
\begin{equation}
(R_{\theta}\otimes R_{\theta^{\prime}})\rho_{AB}(R_{\theta}\otimes
R_{\theta^{\prime}})^{\dagger}=\rho_{AB}~, \label{InvROT}%
\end{equation}
so that they are perfectly transmitted by the bosonic twirling environment%
\begin{equation}
\mathcal{E}_{\theta\theta^{\prime}}(\rho_{AB})=\rho_{AB}~.
\end{equation}
Let us start this search within the set of zero-mean Gaussian states,
therefore completely characterized by their CMs $\mathbf{V}_{AB}$. Finding a
solution of Eq.~(\ref{InvROT}) is therefore equivalent to solving
\begin{equation}
(\mathbf{R}_{\theta}\mathbf{\oplus R}_{\theta^{\prime}})\mathbf{V}%
_{AB}(\mathbf{R}_{\theta}\mathbf{\oplus R}_{\theta^{\prime}})^{T}%
=\mathbf{V}_{AB}~. \label{InvROT2}%
\end{equation}
Depending on the type of environment, i.e., $\theta^{\prime}=\theta$
(correlated rotations)\ or $\theta^{\prime}=-\theta$ (anti-correlated
rotations), we have two different classes of invariant Gaussian states.

Unfortunately, in the case of correlated rotations, the invariant Gaussian
states are separable. In fact, it is easy to check that, for $\theta^{\prime
}=\theta$ and arbitrary $\theta$, the unique solution of Eq. (\ref{InvROT2})
is given by the quasi-normal form%
\begin{equation}
\mathbf{V}_{AB}:=\left(
\begin{array}
[c]{cc}%
\mathbf{A} & \mathbf{C}\\
\mathbf{C}^{T} & \mathbf{B}%
\end{array}
\right)  =\left(
\begin{array}
[c]{cccc}%
\alpha &  & \omega & \varphi\\
& \alpha & -\varphi & \omega\\
\omega & -\varphi & \beta & \\
\varphi & \omega &  & \beta
\end{array}
\right)  ~, \label{CM_appfinite}%
\end{equation}
with $\alpha,\beta\geq1$ and $\omega,\delta$ are real numbers (which must
satisfy a set of bona-fide conditions in order to make the previous matrix a
quantum CM~\cite{TwomodePRA}). Then, it is easy to check that the previous CM
can describe separable Gaussian states only. See Appendix~\ref{invGAUSapp} to
see how to derive Eq.~(\ref{CM_appfinite}) and check its separability properties.

Thus, despite there are two-mode Gaussian states $\rho_{AB}$ which are
invariant under correlated rotations ($\theta^{\prime}=\theta$), these states
must be separable. This negative result can be generalized: No entangled
Gaussian state is invariant under twirlings of the form $U\otimes U $, with
$U$ Gaussian unitary (apart from the trivial case $U=\pm I$). For a simple
proof see Appendix~\ref{APPgtwirl}.

Luckily, the scenario is completely different when we consider the other type
of environment. We can easily find entangled Gaussian states which are
invariant under anti-correlated rotations. In fact, we can easily check that,
for $\theta^{\prime}=-\theta$ and arbitrary $\theta$, a possible solution of
Eq. (\ref{InvROT2}) is given by the CM%
\begin{equation}
\mathbf{V}_{AB}(\mu):=\left(
\begin{array}
[c]{cc}%
\mu\mathbf{I} & \mu^{\prime}\mathbf{Z}\\
\mu^{\prime}\mathbf{Z} & \mu\mathbf{I}%
\end{array}
\right)  ~, \label{EPRcmMU}%
\end{equation}
where%
\begin{equation}
\mu\geq1,~\mu^{\prime}:=\sqrt{\mu^{2}-1},
\end{equation}
and%
\begin{equation}
\mathbf{I}=\left(
\begin{array}
[c]{cc}%
1 & \\
& 1
\end{array}
\right)  ,~\mathbf{Z}=\left(
\begin{array}
[c]{cc}%
1 & \\
& -1
\end{array}
\right)  .
\end{equation}
This is the CM of a two-mode squeezed vacuum state $\rho_{AB}(\mu)$, also
known as EPR state~\cite{RMP}. Thus, in the presence of the anti-correlated
environment $\mathcal{E}_{\theta(-\theta)}$, despite Charlie cannot share any
entanglement with Alice or Bob (one-mode entanglement-breaking), he is still
able to distribute them entanglement by transmitting EPR states perfectly
(two-mode entanglement-preserving).

More generally, we can extend the previous analysis and include non-Gaussian
input states in Charlie's hands. One possible choice is the
continuous-variable version of the Werner state~\cite{cvWERNER}%
\begin{equation}
\rho_{AB}(p,\mu)=p~\rho_{AB}(\mu)+(1-p)~[\rho_{A}(\mu)\otimes\rho_{B}(\mu)]
\label{CVwerner}%
\end{equation}
which corresponds to a mix, with probability $p$, of an EPR state $\rho
_{AB}(\mu)$ and a tensor-product of two identical single-mode thermal states,
$\rho_{A}(\mu)$ and $\rho_{B}(\mu)$, with CMs $\mathbf{V}_{A}(\mu
)=\mathbf{V}_{B}(\mu)=\mu\mathbf{I}$. This state is known to be entangled
for~\cite{cvWERNER}%
\begin{equation}
p>p(\bar{n}):=\frac{1}{1+2\coth(2~\text{\textrm{arcsinh}}\sqrt{\bar{n}})}~,
\end{equation}
where $\bar{n}=(\mu-1)/2$ is the mean number of thermal photons in each mode
[for large $\bar{n}$, we have $p(\bar{n})\rightarrow1/3$, which reminds the
threshold valid for qubit Werner states].

Since thermal states are invariant under phase-space rotations, the convex
combination of Eq.~(\ref{CVwerner}) is invariant under the action of
anti-correlated rotations $\theta^{\prime}=-\theta$, therefore representing a
fixed point of $\mathcal{E}_{\theta(-\theta)}$. As a result, any
continuous-variable Werner state $\rho_{AB}(p,\mu)$ with $p>p(\bar{n})$ can be
used to perfectly transmit entanglement to Alice and Bob, despite the
environment $\mathcal{E}_{\theta(-\theta)}$ being one-mode entanglement-breaking.

\section{Entanglement restoration in correlated-noise Gaussian environments
\label{SECgaussMAIN}}

In the previous Sec.~\ref{SECtwirl}, we have studied the distribution of
entanglement in twirling channels, considering quantum systems with Hilbert
spaces of any dimension, both finite and infinite. We have shown the
conditions under which these environments are simultaneously one-system
entanglement-breaking and two-system entanglement-preserving. This peculiar
situation is possible due to the existence of invariant subspaces of entangled
states (Werner states and isotropic states). In this case, entanglement
preservation is induced by the injection of purely-classical correlations in
the travelling systems.

In this section, we consider a rather different scenario. We construct a
correlated-noise Gaussian environment which generalizes the standard model of
memoryless thermal-lossy environment to include cross-correlations between two
travelling modes. This is done by using two beam-splitters which mix the
travelling modes, $A$ and $B$, with two environmental modes, $E_{1}$ and
$E_{2}$, prepared in a correlated Gaussian state $\rho_{E_{1}E_{2}}$ (in
particular, it can be chosen to be separable). Each single beam-splitter
introduces losses and thermal noise such to realize an entanglement-breaking
channel. For this reason, Charlie is not able to share any entanglement with
Alice or Bob. Despite this, the noise-correlations enable Charlie to
distribute entanglement to Alice and Bob. As we will show, an input EPR state
$\rho_{AB}$\ with sufficiently-high EPR\ correlations is able to generate an
output state which is entangled and even distillable by Alice and Bob.

This is remarkable considering that: (i) For these environments, there is no
decoherence-free\ subspace of entangled states (which means that entanglement
cannot be preserved); and (ii) Entanglement distribution can be reactivated by
the injection of separable correlations (so that lost entanglement is not
replaced by other entanglement, i.e., coming from the environment).

In detail, this section is structured as follows. In Sec.~\ref{SECgauss}, we
characterize our basic model of correlated-noise Gaussian environment,
identifying the physical conditions under which it is separable or entangled
(Sec.~\ref{sepCMsubsec}), and discussing specific kinds of quadrature
correlations, that we call symmetric or asymmetric (Sec.~\ref{subSECspecific}%
). In Sec.~\ref{SECdirect}, we study the distribution of entanglement.
Assuming the condition of one-mode entanglement-breaking, we show how
entanglement distribution can be restored by the environmental correlations
(in particular, separable correlations) identifying the regime where remote
entanglement is also distillable. We show that entanglement restoration is a
threshold process starting only after a critical amount of correlations is
present in the environment.

In particular, in Secs.~\ref{subSYM} and~\ref{subASYM}, we specify this study
to environments with symmetric and asymmetric correlations, analyzing the
process of entanglement restoration in terms of the injected classical and
quantum correlations. Here we compute the critical number of correlation bits
(cbits and dbits) which are needed for entanglement restoration. Once
reactivated, the number of ebits remotely restored is increasing in the number
of correlation bits which are injected. This monotonic behavior holds true
when we compare environments with the same types of correlations (symmetric or
asymmetric) but fails to be true in the general case as explicitly discussed
in Sec.~\ref{nonMONO}.

Finally, in Sec.~\ref{subEPRevolution}, we also analyze the evolution of the
EPR\ correlations in this peculiar process. General notions and technical
details on bosonic systems and Gaussian states can be found in
Appendix~\ref{APPbosonic}.

\subsection{Characterization of Gaussian environments with correlated noise
\label{SECgauss}}

In this section, we introduce and fully characterize our model of
correlated-noise Gaussian environment, which represents the simplest and most
direct generalization of the standard memoryless Gaussian environment with
losses and thermal noise.

\begin{figure}[ptbh]
\vspace{-2.4cm}
\par
\begin{center}
\includegraphics[width=0.65\textwidth] {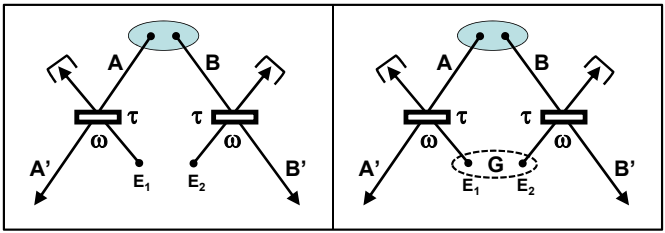}
\end{center}
\par
\vspace{-2.5cm}\caption{Gaussian environments.\textit{\ Left}. Memoryless
Gaussian environment with losses $\tau$ and thermal noise $\omega$.
\textit{Right}. Correlated-noise Gaussian environment, with losses $\tau$,
thermal noise $\omega$ and cross-correlations $\mathbf{G}$. The state of the
environment $\rho_{E_{1}E_{2}}$ can be separable or entangled.}%
\label{ENVschemes}%
\end{figure}

We consider a Gaussian decoherence process which affects two bosonic modes,
$A$ and $B$, in a symmetric way. In order to describe the introduction of
losses and thermal noise, we consider two beam splitters (with the same
transmissivity $\tau$) which combine modes $A$ and $B$ with two environmental
modes, $E_{1}$ and $E_{2}$, respectively. These ancillary modes are prepared
in a zero-mean Gaussian state $\rho_{E_{1}E_{2}}$ which is symmetric under
$E_{1}$-$E_{2}$ permutation. In the standard memoryless model depicted in the
left panel of Fig.~\ref{ENVschemes}, the environment is in a tensor-product
state $\rho_{E_{1}E_{2}}=\rho_{E_{1}}\otimes\rho_{E_{2}}$, meaning that
$E_{1}$ and $E_{2}$ are fully independent. In particular, $\rho_{E_{1}}%
=\rho_{E_{2}}$ is a thermal state with covariance matrix (CM) $\omega
\mathbf{I}$, where the noise variance $\omega=2\bar{n}+1$ quantifies the mean
number of thermal photons $\bar{n}$\ entering the beam splitter. Each
interaction is then equivalent to a lossy channel with transmissivity $\tau$
and thermal noise $\omega$.

In our work, we generalize this Gaussian process to include the presence of
correlations between the environmental modes, as depicted in the right panel
of Fig.~\ref{ENVschemes}. The simplest extension of the model consists\ of
taking the ancillary modes, $E_{1}$ and $E_{2}$, in a zero-mean Gaussian state
$\rho_{E_{1}E_{2}}$ with CM in the symmetric normal form
\begin{equation}
\mathbf{V}_{E_{1}E_{2}}(\omega,g,g^{\prime})=\left(
\begin{array}
[c]{cc}%
\omega\mathbf{I} & \mathbf{G}\\
\mathbf{G} & \omega\mathbf{I}%
\end{array}
\right)  ~, \label{EVE_cmAPP}%
\end{equation}
where $\omega\geq1$ is the thermal noise variance associated with each
ancilla, and the off-diagonal block%
\begin{equation}
\mathbf{G=}\left(
\begin{array}
[c]{cc}%
g & \\
& g^{\prime}%
\end{array}
\right)  \label{Gblock}%
\end{equation}
accounts for the correlations between the ancillas. Such a correlated thermal
state can be separable or entangled (explicit conditions will be given below),
and its quantum discord is always non-zero (apart when $g^{\prime}=g=0$).

It is clear that, when we consider the two interactions $A-E_{1}$ and
$B-E_{2}$ separately, the environmental correlations are washed away. In fact,
by tracing out $E_{2}$, we are left with mode $E_{1}$ in a thermal state
($\mathbf{V}_{E_{1}}=\omega\mathbf{I}$) which is combined with mode $A$ via
the beam-splitter. In other words, we have again a lossy channel with
transmissivity $\tau$ and thermal noise $\omega$. The scenario is identical
for the other mode $B$ when we trace out $E_{1}$. However, when we consider
the joint action of the two environmental modes, the correlation block
$\mathbf{G}$ comes into play and the global dynamics of the two travelling
modes becomes completely different from the standard memoryless scenario.

Before studying the system dynamics and the corresponding evolution of
entanglement, we need to characterize the correlation block $\mathbf{G}$\ more
precisely. In fact, the two correlation parameters, $g$ and $g^{\prime}$,
cannot be completely arbitrary but must satisfy specific physical constraints.
These parameters must vary within ranges which make the CM of
Eq.~(\ref{EVE_cmAPP}) a bona-fide quantum CM~\cite{TwomodePRA}. Given an
arbitrary value of the thermal noise $\omega\geq1$, the correlation parameters
must satisfy the following three bona-fide conditions
\begin{equation}
|g|<\omega,~~~|g^{\prime}|<\omega,~~~\omega^{2}+gg^{\prime}-1\geq
\omega\left\vert g+g^{\prime}\right\vert . \label{CMconstraints}%
\end{equation}
(Details of the proof can be found in Appendix~\ref{APP_sub2modes}).

\subsubsection{Separability properties\label{sepCMsubsec}}

Once the bona-fide conditions for the environment are fully clarified, the
next step is the characterization of its separability properties.

For this aim, we compute the smallest partially-transposed symplectic (PTS)
eigenvalue $\varepsilon_{\text{env}}$ associated with the CM of the
environment\ $\mathbf{V}_{E_{1}E_{2}}$ (for more details on this formalism see
Appendix~\ref{APP_sub2modes}). For Gaussian states, this eigenvalue represents
an entanglement monotone, being fully equivalent to the
log-negativity~\cite{logNEG}. After simple algebra, we get%
\begin{equation}
\varepsilon_{\text{env}}=\sqrt{\omega^{2}-gg^{\prime}-\omega|g-g^{\prime}|}~.
\end{equation}
Provided that the conditions of Eq.~(\ref{CMconstraints}) are satisfied, the
separability condition $\varepsilon_{\text{env}}\geq1$ is equivalent to%
\begin{equation}
\omega^{2}-gg^{\prime}-1\geq\omega|g-g^{\prime}|~. \label{sepCON}%
\end{equation}

The various conditions of bona-fide and separability can be combined together.
An environment of the form~(\ref{EVE_cmAPP}) with thermal noise $\omega\geq1$
is bona-fide and separable when the correlation parameters $g$ and $g^{\prime
}$ satisfy%
\begin{equation}
|g|<\omega,~~~|g^{\prime}|<\omega,~~~\omega^{2}-1\geq\max\{\Gamma_{-}%
,\Gamma_{+}\}~. \label{sepENV}%
\end{equation}
where%
\begin{align}
\Gamma_{-}  &  :=\omega\left\vert g+g^{\prime}\right\vert -gg^{\prime}~,\\
\Gamma_{+}  &  :=\omega\left\vert g-g^{\prime}\right\vert +gg^{\prime}~.
\end{align}
By contrast, it is bona-fide and entangled when%
\begin{equation}
|g|<\omega,~~~|g^{\prime}|<\omega,~~~\Gamma_{-}\leq\omega^{2}-1<\Gamma_{+}~.
\label{entENV}%
\end{equation}

\begin{figure}[ptbh]
\vspace{-0.0cm}
\par
\begin{center}
\includegraphics[width=0.47\textwidth] {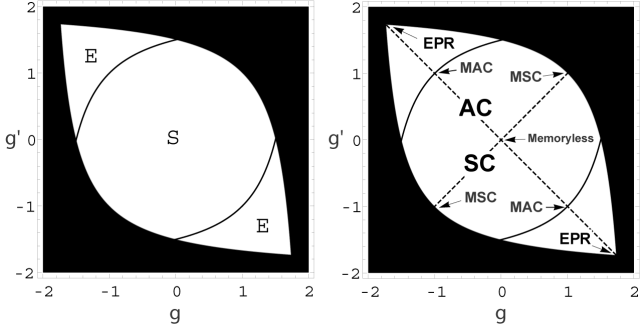}
\end{center}
\par
\vspace{-0.6cm}\caption{\textit{Left.} Correlation plan $(g,g^{\prime})$ for
the Gaussian environment, corresponding to thermal noise $\omega=2$. The black
area identifies forbidden environments (correlations are too strong to be
compatible with quantum mechanics). White area identifies physical
environments, i.e., the subset of points which satisfy the bona-fide
conditions of Eq.~(\ref{CMconstraints}). Within this area, the inner region
labelled by \textquotedblleft S\textquotedblright\ identifies separable
environments [Eq.~(\ref{sepENV}) is satisfied] while the two outer regions
labelled by \textquotedblleft E\textquotedblright\ identify entangled
environments [Eq.~(\ref{entENV}) is satisfied].\textit{\ Right.} Correlation
plan where we display the memoryless environment (origin $g^{\prime}=g=0$), SC
environments (bisector $g^{\prime}=g$) and AC\ environments (bisector
$g^{\prime}=-g$). The MSC environments are the SC environments with maximal
correlations (extremal points on the bisector $g^{\prime}=g$). The EPR
environments are the AC\ environments with maximal correlations (extremal
points on the bisector $g^{\prime}=-g$). Finally, the MAC environments are
those AC environments which are simultaneously separable and maximally
correlated.}%
\label{ENVcha}%
\end{figure}To better clarify the structure of the environment, we provide a
numerical example in Fig.~\ref{ENVcha}. In the left panel of this figure, we
consider the \textit{correlation plan} which is spanned by the two parameters
$g$ and $g^{\prime}$. For a given value of the thermal noise $\omega$, we
identify the subset of points which satisfy the bona-fide conditions of
Eq.~(\ref{CMconstraints}). This subset corresponds to the white area in the
figure. Within this area, we then characterize the regions which correspond to
separable environments (area labelled by S) and entangled environments (areas
labelled by E).

\subsubsection{Symmetric and asymmetric noise
correlations\label{subSECspecific}}

Here we specify our model of Gaussian environment to particular cases with
specific correlation properties. Before discussing these cases, note that the
memoryless Gaussian environment corresponds to the origin of the correlation
plan ($g^{\prime}=g=0$), as also depicted in the right panel of
Fig.~\ref{ENVcha}. In particular, the memoryless environment represents the
unique physical solution for $\omega=1$ (vacuum noise), in correspondence to
which all the correlation plan collapses into its origin. In the remainder of
this section we implicitly take $\omega>1$, therefore excluding the trivial
case of a singular correlation plan.

The first type of specifically-correlated environment is called
\textquotedblleft symmetrically correlated\textquotedblright\ (SC). This
corresponds to taking $g^{\prime}=g$, which means that positions and momenta
of the two modes, $E_{1}$ and $E_{2}$, are correlated exactly in the same way.
On the correlation plan, these environments are points on the bisectors of the
first and third quadrants (see the right panel of Fig.~\ref{ENVcha}).

It is easy to check that the bona-fide conditions of Eq.~(\ref{CMconstraints})
simplify to the single inequality%
\begin{equation}
|g|\leq\omega-1~, \label{qSymmCON}%
\end{equation}
with maximal symmetrical correlations (MSC) achieved at $g=\omega-1$ and
$g=1-\omega$, which are the two extremal points shown in the figure. It is
important to note that SC environments are always separable, since
Eq.~(\ref{sepCON}) becomes $|g|\leq\sqrt{\omega^{2}-1}$, which is always
satisfied by Eq.~(\ref{qSymmCON}). Then, they are also mixed, with von Neumann
entropy equal to
\begin{equation}
S_{\text{SC}}(\omega,g)=h(\omega+g)+h(\omega-g)~,
\end{equation}
where~\footnote{In all the paper, \textquotedblleft$\log$\textquotedblright%
\ is intended to be base $2$, so that the corresponding quantities are
measured in bits. We write \textquotedblleft$\ln$\textquotedblright\ for the
natural base $e$.}%
\begin{equation}
h(x):=\frac{x+1}{2}\log\left(  \frac{x+1}{2}\right)  -\frac{x-1}{2}\log\left(
\frac{x-1}{2}\right)  . \label{hVonNEUMANN}%
\end{equation}

Given an arbitrary SC environment, we can investigate the nature of its
separable correlations in terms of classical correlations (measured in
classical bits or \textquotedblleft cbits\textquotedblright) and Gaussian
discord (measured in discordant bits or \textquotedblleft
dbits\textquotedblright). Using the formulas of Ref.~\cite{GerryD} (see also
Appendix~\ref{APP_subDISCORD}) we compute the amount of classical correlations
which is here equal to%
\begin{equation}
C=h(\omega)-h\left(  \omega-\frac{g^{2}}{\omega+1}\right)  , \label{Csym}%
\end{equation}
and the amount of Gaussian discord $D$ which is given by%
\begin{equation}
D_{\text{SC}}(\omega,g)=h(\omega)+h\left(  \omega-\frac{g^{2}}{\omega
+1}\right)  -S_{\text{SC}}(\omega,g).
\end{equation}
The bits of total correlations are quantified by the quantum mutual
information $I=C+D$, here equal to%
\begin{equation}
I_{\text{SC}}(\omega,g)=2h(\omega)-S_{\text{SC}}(\omega,g)~.
\end{equation}
It is important to note that Gaussian discord $D$ is an upper bound to the
actual discord, even if they are conjectured to
coincide~\cite{DisConj1,DisConj2}. Correspondingly, the measure for classical
correlations which is here considered $C=I-D$ represents a lower bound to the
actual amount of classical correlations.

As expected, both types of correlations, $C$ and $D$, are increasing in
$\left\vert g\right\vert $ at any fixed value of thermal noise $\omega$.
Despite increasing, Gaussian discord must be bounded by $D\leq1$, which is
compatible with the fact that SC\ environments are always separable ($D>1$ is
a sufficient condition for entanglement~\cite{GerryD}). Finally, one can
check~\cite{StateDIS} that MSC environments\ have maximal Gaussian discord not
only between the SC environments but, more generally, among all the separable
Gaussian environments, represented by the \textquotedblleft
S\textquotedblright\ region in Fig.~\ref{ENVcha}.

The second type of environment, that we call\textit{\ \textquotedblleft%
}asymmetrically correlated\textquotedblright\ (AC), corresponds to the
condition $g^{\prime}=-g$. This means that positions and momenta have opposite
correlations, i.e., if positions are correlated (anticorrelated), then momenta
are anticorrelated (correlated). On the correlation plan, these environments
are represented by the points lying on the bisectors of the second and fourth
quadrants (see right panel of Fig.~\ref{ENVcha}).

In this case, the bona-fide conditions of Eq.~(\ref{CMconstraints}) simplify
to the inequality%
\begin{equation}
|g|\leq\sqrt{\omega^{2}-1}~. \label{qAsymmCON}%
\end{equation}
For maximal correlations $|g|=\sqrt{\omega^{2}-1}$ we have an EPR state, which
is pure and maximally entangled. Depending on the sign of $g$, we have two
different EPR environments: The positive EPR environment ($g=\sqrt{\omega
^{2}-1}$) with positions correlated, and the negative EPR environment
($g=-\sqrt{\omega^{2}-1}$) with positions anticorrelated. Apart from these
extremal cases, AC environments are mixed with entropy%
\begin{equation}
S_{\text{AC}}(\omega,g)=2h\left(  \sqrt{\omega^{2}-g^{2}}\right)  ~.
\end{equation}

In general, we note that AC environments can be separable or entangled. By
using Eq.~(\ref{sepCON}), we see that they are separable for
\begin{equation}
|g|\leq\omega-1~,
\end{equation}
while they are entangled for stronger correlations
\begin{equation}
\omega-1<|g|\leq\sqrt{\omega^{2}-1}~. \label{QcorrelatedENV}%
\end{equation}
We call maximally-separable AC environments (MACs) those AC environments with
maximal correlations but still separable. They are characterized by the border
condition $|g|=\omega-1$, and they correspond to the two intersection points
shown in the right panel of Fig.~\ref{ENVcha}.

As before, we can easily quantify the amounts of classical correlations $C$
and Gaussian discord $D$, both increasing functions in $\left\vert
g\right\vert $. For AC environments, these quantities are respectively given
by Eq.~(\ref{Csym}) and%
\begin{equation}
D_{\text{AC}}(\omega,g)=h(\omega)+h\left(  \omega-\frac{g^{2}}{\omega
+1}\right)  -S_{\text{AC}}(\omega,g)~.
\end{equation}
The quantum mutual information is now equal to%
\[
I_{\text{AC}}(\omega,g)=2h(\omega)-S_{\text{AC}}(\omega,g)~.
\]

The correlation properties of the AC environments can be easily compared with
those of the previous SC environments. For fixed values of $\omega$ and $g$
with $\left\vert g\right\vert \leq\omega-1$ (separable environments), AC and
SC environments have identical classical correlations but different Gaussian
discord, i.e.,%
\begin{equation}
\delta D:=D_{\text{SC}}(\omega,g)-D_{\text{AC}}(\omega,g)>0~.
\end{equation}
This means that they also have different amounts of total correlations
\begin{equation}
\delta I:=I_{\text{SC}}(\omega,g)-I_{\text{AC}}(\omega,g)>0~.
\end{equation}
As clear from the previous formulas, this is simply a consequence of the fact
that AC environments are more entropic than SC\ environments, i.e.,
\begin{equation}
\delta S:=S_{\text{SC}}(\omega,g)-S_{\text{AC}}(\omega,g)<0~.
\end{equation}
The difference in entropy directly quantifies the difference in the
correlations between these environments%
\begin{equation}
\delta D=\delta I=-\delta S~.
\end{equation}

However, contrary to the SC environments, the AC environments can also become
entangled. When the correlation parameter $g$ is chosen in the entanglement
regime of Eq.~(\ref{QcorrelatedENV}), classical correlations and Gaussian
discord become larger (in particular, $D$ may exceed $1$). The optimum is
reached at the two EPR states whose classical and quantum correlations are
identical and maximal in the entire correlation plan. In particular, we have
$C_{\text{EPR}}=D_{\text{EPR}}=h(\omega)$ which is just the entropy of
entanglement of the EPR state.

\subsection{Distribution and distillation of entanglement in correlated-noise
Gaussian environments\label{SECdirect}}

\begin{figure}[ptbh]
\vspace{-2.3cm}
\par
\begin{center}
\includegraphics[width=0.60\textwidth] {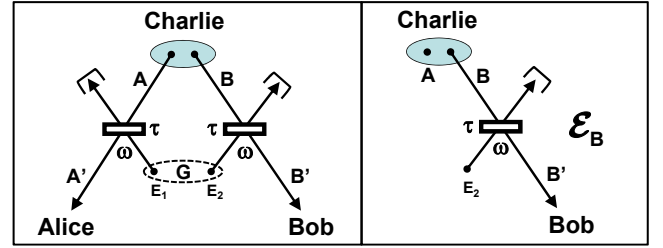}
\end{center}
\par
\vspace{-2.4cm}\caption{Scenarios for entanglement distribution.
\textit{Left}.~Charlie has two modes $A$ and $B$ prepared in an EPR state
$\rho_{AB}$. In order to distribute entanglement to the remote parties,
Charlie transmits the two modes through the correlated Gaussian environment,
characterized by transmissivity $\tau$, thermal noise $\omega$ and
correlations $\mathbf{G}$. \textit{Right}.~Charlie aims to share entanglement
with a remote party (e.g., Bob). He then keeps mode $A$ while sending mode $B$
through the lossy channel $\mathcal{E}_{B}$.}%
\label{twomodeEB}%
\end{figure}

Let us study the propagation of entanglement in a correlated-noise Gaussian
environment. Suppose that Charlie has an entanglement source described by an
EPR\ state $\rho_{AB}$ with CM $\mathbf{V}_{AB}(\mu)$ given in
Eq.~(\ref{EPRcmMU}).\textbf{\ }We consider the different scenarios depicted in
the two panels of Fig.~\ref{twomodeEB}. Charlie may attempt to distribute
entanglement to Alice and Bob as shown in the left panel of
Fig.~\ref{twomodeEB}, or he may try to share entanglement with one of the
remote parties, e.g., Bob, as shown in the right panel of Fig.~\ref{twomodeEB}
(by symmetry, our derivation is the same if we consider Alice in the place of Bob).

Let us start from the scenario where Charlie aims to share entanglement with
Bob, which means that he keeps mode $A$ while sending mode $B$. The action of
the environment is therefore reduced to $\mathcal{I}_{A}\otimes\mathcal{E}%
_{B}$, where $\mathcal{E}_{B}$ is a lossy channel with transmissivity $\tau$
and thermal noise $\omega$. It is easy to check~\cite{NoteCM1} that the output
state $\rho_{AB^{\prime}}$, shared by Charlie and Bob, is Gaussian with zero
mean and CM%
\begin{equation}
\mathbf{V}_{AB^{\prime}}=\left(
\begin{array}
[c]{cc}%
\mu\mathbf{I} & \mu^{\prime}\sqrt{\tau}\mathbf{Z}\\
\mu^{\prime}\sqrt{\tau}\mathbf{Z} & x\mathbf{I}%
\end{array}
\right)  , \label{ABpCM}%
\end{equation}
where
\begin{equation}
x:=\tau\mu+(1-\tau)\omega~.
\end{equation}

We can compute closed analytical formulas in the limit of large $\mu$, i.e.,
large input entanglement. In this case, the entanglement of the output state
$\rho_{AB^{\prime}}$ is quantified by the PTS\ eigenvalue%
\begin{equation}
\varepsilon=\frac{1-\tau}{1+\tau}\omega~.
\end{equation}
The entanglement-breaking condition corresponds to the separability condition
$\varepsilon\geq1$, which provides%
\begin{equation}
\omega\geq\frac{1+\tau}{1-\tau}:=\omega_{\text{EB}}~, \label{EBcond}%
\end{equation}
or, equivalently, $\bar{n}\geq\tau/(1-\tau)$ for the mean number of thermal photons.

Despite the entanglement-breaking condition of Eq.~(\ref{EBcond}) being
derived for an EPR\ input, it is indeed valid for any input state. In other
words, a lossy channel $\mathcal{E}_{B}$\ with transmissivity $\tau$ and
thermal noise $\omega\geq\omega_{\text{EB}}$ destroys the entanglement of any
input state $\rho_{AB}$. As one can check, Eq.~(\ref{EBcond}) corresponds
exactly to the entanglement-breaking condition for lossy channels derived in
Ref.~\cite{HolevoEB}. In particular, the threshold condition $\omega
=\omega_{\text{EB}}$ is sufficient to guarantee one-mode
entanglement-breaking, i.e., the impossibility for Charlie to share any
entanglement with a remote party.

Now remember our main question: Suppose that Charlie cannot share any
entanglement with Alice or Bob (one-mode entanglement-breaking), can he still
distribute entanglement to them? In other words, suppose that the
correlated-noise Gaussian environment has transmissivity $\tau$ and thermal
noise $\omega=\omega_{\text{EB}}$, so that the individual lossy channels
$\mathcal{E}_{A}$\ (Charlie$\rightarrow$Alice)\ and $\mathcal{E}_{B}$
(Charlie$\rightarrow$Bob) are entanglement-breaking. Is it still possible to
use the joint channel $\mathcal{E}_{AB}$ to distribute entanglement to the
remote parties? In the following, we explicitly reply to this question. In
particular, we will show that entanglement can be distributed by separable
environments and can be large enough to be distilled by one-way distillation protocols.

Let us derive the general evolution of the two modes $A$ and $B$ under the
action of the joint environment, as depicted in the left panel of
Fig.~\ref{twomodeEB}. Since the input EPR\ state $\rho_{AB}$ is Gaussian and
the environmental state $\rho_{E_{1}E_{2}}$ is Gaussian, the output state
$\rho_{A^{\prime}B^{\prime}}$ is also Gaussian. This state has zero mean and
CM given by%
\begin{equation}
\mathbf{V}_{A^{\prime}B^{\prime}}=\tau\mathbf{V}_{AB}+(1-\tau)\mathbf{V}%
_{E_{1}E_{2}}=\left(
\begin{array}
[c]{cc}%
x\mathbf{I} & \mathbf{H}\\
\mathbf{H} & x\mathbf{I}%
\end{array}
\right)  ~,
\end{equation}
where%
\begin{equation}
\mathbf{H}:=\tau\mu^{\prime}\mathbf{Z}+(1-\tau)\mathbf{G}~.
\end{equation}
For large input entanglement $\mu\gg1$, we can easily derive the symplectic
spectrum of the output state%
\begin{equation}
\nu_{\pm}=\sqrt{\left(  2\omega+g^{\prime}-g\pm|g+g^{\prime}|\right)
(1-\tau)\tau\mu}~,
\end{equation}
and its smallest PTS\ eigenvalue%
\begin{equation}
\varepsilon=(1-\tau)\sqrt{(\omega-g)(\omega+g^{\prime})}~. \label{epsMAIN}%
\end{equation}
The latter eigenvalue determines the remote entanglement distributed to Alice
and Bob, equivalently quantified by the log-negativity%
\begin{equation}
\mathcal{N}=\max\{0,-\log\varepsilon\}~, \label{logNEGmain}%
\end{equation}
which is measured in entanglement bits (ebits). In particular, $\mathcal{N}$
represents an upper bound to the mean number of ebits which are distillable
per output copy $\rho_{A^{\prime}B^{\prime}}$.

In the same limit, we can also compute the coherent information $I(A\rangle
B)$ between the two remote parties, which provides a lower bound to the mean
number of ebits per copy that can be distilled using one-way distillation
protocols, i.e., protocols based on local operations and one-way classical
communication (see Appendix~\ref{APP_sub2modes} for more details). This is
clearly a lower bound since one-way distillability implies two-way
distillability, where both forward and backward communication are generally
exploited. In the remainder of the paper, by distillability we implicitly mean
the sufficient condition of one-way distillability.

For large input entanglement $\mu\gg1$, we can write
\begin{equation}
I(A\rangle B)=-\log(e\varepsilon)~, \label{coheDIR}%
\end{equation}
which is derived by expanding the coherent information for diverging spectra
(see Appendix~\ref{APP_sub2modes}). Thus, remote entanglement is distributed
for $\varepsilon<1$, according to Eq.~(\ref{logNEGmain}), and it is
furthermore distillable for $\varepsilon<e^{-1}$, according to
Eq.~(\ref{coheDIR}).

\begin{figure}[h]
\vspace{-0.0cm}
\par
\begin{center}
\includegraphics[width=0.48\textwidth] {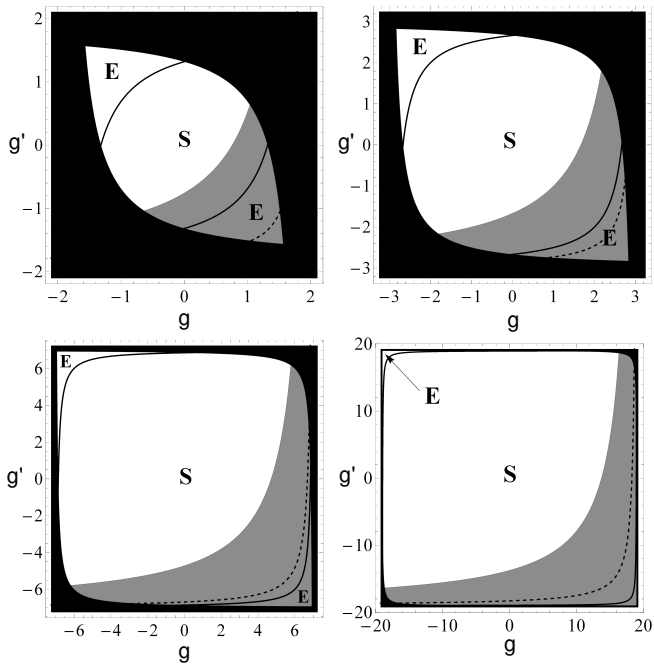}
\end{center}
\par
\vspace{-0.4cm}\caption{Analysis of the remote entanglement $\varepsilon$ on
the correlation plan $(g,g^{\prime})$ for different values of the
transmissivity $\tau=0.3$, $0.5$, $0.75$, and $0.9$ (from top left to bottom
right). Corresponding values of the thermal noise are determined by the
condition of one-mode entanglement-breaking $\omega=\omega_{\text{EB}}$. In
each inset, the non-black area identifies the set of physical environments,
which are divided into separable (S) and entangled (E) environments by the
solid lines. The gray region is the reactivation area\ and identifies those
environments for which Charlie is able to distribute entanglement to Alice and
Bob ($\varepsilon<1$). Within the reactivation area, the points below the
dashed curve are those environments for which the distributed entanglement is
also distillable ($\varepsilon<e^{-1}$).}%
\label{dirTOT}%
\end{figure}

Now suppose that the Gaussian environment has thermal noise $\omega
=\omega_{\text{EB}}$, so that it is one-mode entanglement-breaking. Replacing
in Eq.~(\ref{epsMAIN}), we can write
\begin{align}
\varepsilon &  =\sqrt{[1+\tau-(1-\tau)g][1+\tau+(1-\tau)g^{\prime}%
]}\nonumber\\
&  :=\varepsilon(\tau,g,g^{\prime}) \label{EBentEXP}%
\end{align}
Answering our previous question corresponds to finding environmental
parameters $\tau$, $g$ and $g^{\prime}$, for which\ $\varepsilon$ is
sufficiently low: For a given value of the transmissivity $\tau$, we look for
regions in the correlation plan $(g,g^{\prime})$ where $\varepsilon<1$ (remote
entanglement is distributed) and, more strongly, $\varepsilon<e^{-1}$ (remote
entanglement is distillable). This is done in Fig.~\ref{dirTOT} for several
numerical values of the transmissivity.

In Fig.~\ref{dirTOT}, the environments identified by the gray
\textquotedblleft reactivation area\textquotedblright\ allow Charlie to
distribute entanglement to Alice and Bob ($\varepsilon<1$), despite being
impossible for him to share any entanglement with the remote parties. In other
words, these environments allow two-mode entanglement-distribution, despite
being one-mode entanglement-breaking. Furthermore, we can identify
sufficiently-correlated environments for which the entanglement remotely
distributed is also distillable ($\varepsilon<e^{-1}$). From Fig.~\ref{dirTOT}%
, it is evident that entanglement restoration is a threshold process occurring
only after a critical amount of correlations is injected by the environment.
In particular, the critical values of the correlation parameters, $g$ and
$g^{\prime}$, correspond to the border of the reactivation area.

The most remarkable feature in Fig.~\ref{dirTOT} is represented by the
presence of separable environments in the reactivation area. In other words,
there are separable Gaussian environments which contain enough correlations to
restore the distribution of entanglement to Alice and Bob. Furthermore, for
sufficiently-high transmissivities and correlations, these environments enable
Charlie to distribute a distillable amount of entanglement. For instance, see
the bottom right panel of Fig.~\ref{dirTOT}, where separable environments are predominant.

Thus, we have just proven that a distillable fraction of the large input
entanglement can be remotely restored by the presence of separable
correlations in the Gaussian environment, which are therefore sufficiently
strong to break the mechanism of entanglement-breaking induced by the thermal
noise. As we have already said, achieving the simultaneous conditions of
one-mode entanglement-breaking and two-mode entanglement-distribution is here
surprising, since there is no decoherence-free subspace which can preserve the
input entanglement and there is no entanglement coming from the environment
which can replace the amount lost in the transmission.

By contrast, the conditions of one-mode entanglement-breaking and two-mode
entanglement-distribution can be trivially achieved in entangled environments.
For instance, we may consider two beam-splitters with zero transmissivity, so
that Charlie's state is completely reflected into the environment and the
entangled state of the environment is reflected to Alice and Bob. This
scenario is certainly one-mode entanglement-breaking, since Charlie has no
chance of sharing part of his state with Alice or Bob. At the same time,
two-mode entanglement-distribution is realized, since the loss of Charlie's
initial entanglement is just replaced by the injection of entanglement from
the environment.

\subsubsection{Distribution and distillation in SC\ environments\label{subSYM}%
}

Here we specify the previous analysis to SC environments ($g^{\prime}=g$). We
consider these separable environments at the thermal threshold $\omega
=\omega_{\text{EB}}$, so that they are one-mode entanglement-breaking. In the
limit of large $\mu$, we derive the parameter regimes for which remote
entanglement can be distributed by the two-mode transmission and, more
strongly, result into distillable entanglement. In particular, we provide
simple threshold conditions for the correlation parameter $g$. Finally, we
study the relations between remote entanglement and the various types of
separable correlations (quantum and classical) present in these environments,
quantifying the critical number of correlation bits which are needed for
entanglement restoration.

From Eq.~(\ref{EBentEXP}), we see that Alice and Bob's remote entanglement can
be quantified by%
\begin{equation}
\varepsilon_{\text{SC}}=\varepsilon(\tau,g,g)=\sqrt{(1+\tau)^{2}-(1-\tau
)^{2}g^{2}}. \label{epsSC}%
\end{equation}
Since $\varepsilon_{\text{SC}}$ does not depend on the sign of $g$, its study
can be reduced to SC environments with positive $g$. It is easy to check that
$\varepsilon_{\text{SC}}$ takes its optimal (minimum) value%
\begin{equation}
\varepsilon_{\text{MSC}}=\sqrt{(1-\tau)(1+3\tau)}~, \label{EPSmsc}%
\end{equation}
when the environment has maximal correlations (MSC environment), with
correlation parameter%
\begin{equation}
g_{\text{MSC}}=\omega_{\text{EB}}-1=\frac{2\tau}{1-\tau}~. \label{gMSC}%
\end{equation}

\begin{figure}[ptbh]
\vspace{-2.6cm}
\par
\begin{center}
\includegraphics[width=0.56\textwidth] {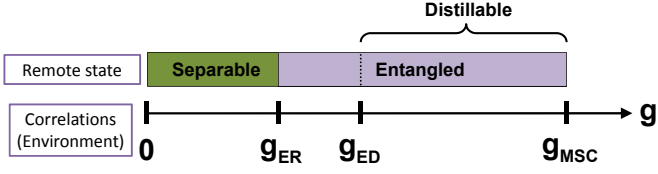}
\end{center}
\par
\vspace{-2.8cm}\caption{Scheme showing the separability properties of Alice
and Bob's remote state in terms of the correlation parameter $g$ of the
SC\ environment (which is always separable). Here we consider the regime of
high transmissivity $\tau\gtrsim0.96$, where remote entanglement can be
distilled for sufficiently high values of the correlation parameter
($g>g_{\text{ED}}$).}%
\label{linea0}%
\end{figure}

From Eq.~(\ref{epsSC}), we can see that remote entanglement is restored
($\varepsilon_{\text{SC}}<1$) for transmissivities $\tau>2/3$ and values of
the correlation parameter $g_{\text{ER}}<g\leq g_{\text{MSC}}$, where
$g_{\text{ER}}$ is the entanglement-restoration threshold%
\begin{equation}
g_{\text{ER}}=\frac{\sqrt{\tau(\tau+2)}}{1-\tau}~.
\end{equation}
More strongly, remote entanglement is distillable ($\varepsilon_{\text{SC}%
}<e^{-1}$) for $\tau\gtrsim0.96$ and $g_{\text{ED}}<g\leq g_{\text{MSC}}$,
where $g_{\text{ED}}$ is the entanglement-distillation threshold%
\begin{equation}
g_{\text{ED}}:=\frac{\sqrt{e^{2}(1+\tau)^{2}-1}}{e(1-\tau)}\geq g_{\text{ER}%
}~.
\end{equation}
See Fig.~\ref{linea0}\ for a pictorial representation. 

\begin{figure}[ptbh]
\vspace{-0.1cm}
\par
\begin{center}
\includegraphics[width=0.36\textwidth] {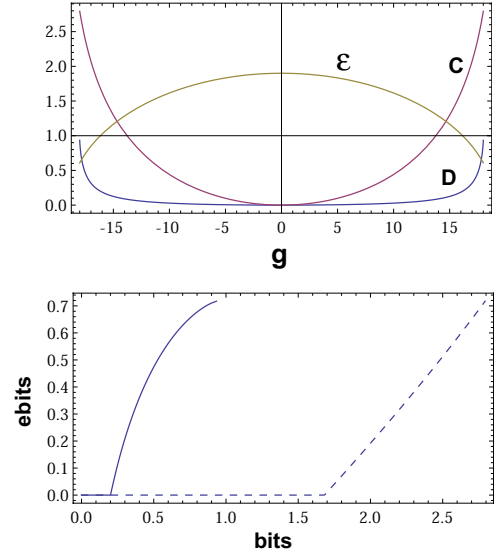}
\end{center}
\par
\vspace{-0.6cm}\caption{\textit{Top panel}. PTS\ eigenvalue $\varepsilon
_{\text{SC}}$ (dimensionless units), Gaussian discord $D$ (dbits) and
classical correlations $C$ (cbits) are plotted versus the correlation
parameter $g$ of the SC environment (note that we can restrict our analysis to
$g>0$ by symmetry). Remote entanglement is generated ($\varepsilon_{\text{SC}%
}<1$) for sufficiently high values of $D$ and $C$, and it is optimal at the
border (MSC environments), where $D$ and $C$ are maximal. In this numerical
example, transmission is $\tau=0.9$ and thermal noise is $\omega
=\omega_{\text{EB}}=$ $19$ (one-mode entanglement-breaking). \textit{Bottom
panel}. Remote entanglement, as quantified by the log-negativity (ebits), is
plotted versus the Gaussian discord (dbits, solid line) and the classical
correlations (cbits, dashed line). Remote entanglement starts to be restored
only after a critical number of correlation bits are injected. Then, it is
monotonically increasing. Numerical parameters are identical as before
($\tau=0.9$ and $\omega=$ $19$).}%
\label{discordSC1}%
\end{figure}

As discussed in Sec.~\ref{subSECspecific}, we can easily quantify the
separable correlations of the SC environments in terms of classical
correlations $C$ and Gaussian discord $D$. These environmental quantities can
be connected with the amount of entanglement which is restored at the remote
stations under the usual conditions one-mode entanglement-breaking
($\omega=\omega_{\text{EB}}$) and large input entanglement ($\mu\gg1$). As we
can see from the top panel of Fig.~\ref{discordSC1}, remote entanglement can
be restored ($\varepsilon_{\text{SC}}<1$) only after a certain amount of
Gaussian discord $D$\ and classical correlations $C$. Plot refers to the
numerical case of $\tau=0.9$, but the same behavior appears for any other
value of transmissivity $\tau>2/3$.

In the bottom panel of Fig.~\ref{discordSC1},\ we have plotted the number of
ebits remotely restored (as quantified by the log-negativity $\mathcal{N}$)
versus the environmental Gaussian discord $D$ (solid line) and classical
correlations $C$ (dashed line). We can clearly see the threshold behavior of
the entanglement, which starts to be restored only after a certain numbers of
correlation bits (dbits and cbits) are injected by the environment. After
these critical values, the number of remote ebits is monotonically increasing.

\begin{figure}[ptbh]
\vspace{-0.6cm}
\par
\begin{center}
\includegraphics[width=0.43\textwidth] {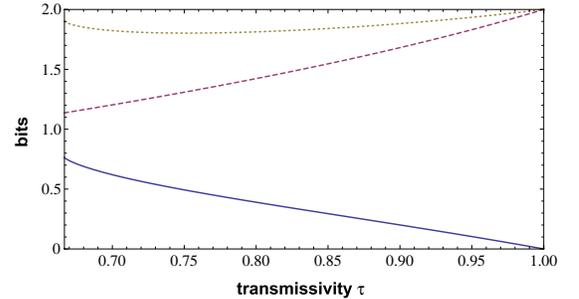}
\end{center}
\par
\vspace{-0.6cm}\caption{For SC environments at any transmissivity $\tau>2/3$,
we plot the critical number of correlation bits after which remote
entanglement starts to be restored. We show the critical number of dbits
(solid), cbits (dashed) and their total (dotted), corresponding to bits of
quantum mutual information.}%
\label{criticalscPIC}%
\end{figure}

Entanglement restoration is a threshold process at any transmissivity
$\tau>2/3$. This is particularly evident from Fig.~\ref{criticalscPIC}, where
we plot the critical numbers of dbits and cbits, after which remote
entanglement is established. For completeness, we also plot the critical
number of bits of total correlations (quantified by the quantum mutual
information) which are needed for the restoration. As we can see, we need less
than 2 bits of total correlations to recover entanglement at any $\tau>2/3$.
Also note that, for extremely low loss ($\tau\rightarrow1$) and high thermal
noise ($\omega_{\text{EB}}\rightarrow+\infty$), restoration is achieved in
correspondence of a negligible amount of Gaussian discord and $2$ bits of
classical correlations. 

\subsubsection{Distribution and distillation in
AC\ environments\label{subASYM}}

Here we repeat the previous analysis for the case of AC environments
($g^{\prime}=-g$), under the same conditions of one-mode entanglement-breaking
($\omega=\omega_{\text{EB}}$) and large input entanglement ($\mu\gg1$). Using
Eq.~(\ref{EBentEXP}), we see that Alice and Bob's remote entanglement is
quantified by%
\begin{equation}
\varepsilon_{\text{AC}}=\varepsilon(\tau,g,-g)=1+\tau-(1-\tau)g~.
\label{EPSepr}%
\end{equation}
We can have $\varepsilon_{\text{AC}}<1$ only for positive values of the
correlation parameter $g$. Such an asymmetry, which is also evident from
Fig.~\ref{dirTOT}, depends on the fact that Charlie's input state has EPR
correlations of the type $\hat{q}_{A}\simeq\hat{q}_{B}$ and $\hat{p}_{A}%
\simeq-\hat{p}_{B}$. These input correlations tend to be preserved by AC
environments with positive $g$ (whose correlations are of the same type) while
they tend to be destroyed by AC environments with negative $g$ (opposite type).

For this reason, in our analysis we only consider positive AC environments
($g>0$) which are separable for $g\leq g_{\text{MAC}}=\omega_{\text{EB}}-1$,
and entangled for%
\begin{equation}
g_{\text{MAC}}<g\leq g_{\text{EPR}}=\sqrt{\omega_{\text{EB}}^{2}-1}%
=\frac{2\sqrt{\tau}}{1-\tau}.
\end{equation}
The optimal restoration of entanglement is achieved by the positive EPR
environment $g=g_{\text{EPR}}$, for which we have $\varepsilon_{\text{EPR}%
}=(1-\sqrt{\tau})^{2}$. One can easily check that this is the global optimum
in the entire correlation plan. Optimal restoration in separable AC
environments is achieved by the positive MAC environment ($g=g_{\text{MAC}}$)
for which
\begin{equation}
\varepsilon_{\text{MAC}}=1-\tau~. \label{EPSmac}%
\end{equation}

\begin{figure}[ptbh]
\vspace{-2.3cm}
\par
\begin{center}
\includegraphics[width=0.53\textwidth] {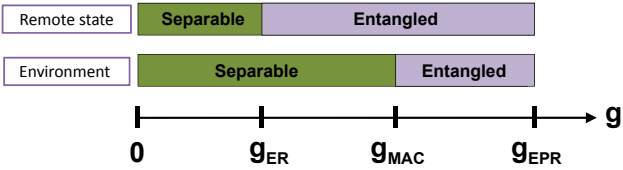}
\end{center}
\par
\vspace{-2.6cm}\caption{Separability properties of the remote state versus
those of the positive AC environments ($g>0$). For $g_{\text{ER}}<g\leq
g_{\text{MAC}}$, there are separable AC\ environments able to restore remote
entanglement.}%
\label{linea1}%
\end{figure}

In general, from Eq.~(\ref{EPSepr}), we see that remote entanglement is
restored at any transmissivity $\tau$ for values of the correlation parameter
$g_{\text{ER}}<g\leq g_{\text{EPR}}$, where the entanglement-restoration
threshold is here equal to%
\begin{equation}
g_{\text{ER}}:=\frac{\tau}{1-\tau}~. \label{gEGasymm}%
\end{equation}
This threshold satisfies $g_{\text{ER}}=g_{\text{MAC}}/2$, which means that
remote entanglement can always be recovered by separable AC environments with
sufficiently high correlations, as pictorially shown in Fig.~\ref{linea1}.

Thus, at any transmissivity $\tau$, there always exist separable
AC\ environments able to reactivate the entanglement distribution. In
particular, this is true for $\tau\simeq0$ (and therefore $\omega_{\text{EB}%
}\simeq1$). In fact, by expanding at the leading order in $\tau$, we get%
\begin{equation}
g_{\text{ER}}\simeq\tau~,~g_{\text{MAC}}\simeq2\tau~,~g_{\text{EPR}}%
\simeq2\sqrt{\tau}~,
\end{equation}
which means that we can always find a correlation parameter $g\simeq0$ such
that $g_{\text{ER}}<g\leq g_{\text{MAC}}$. Remarkably, in conditions of
extreme losses, where entanglement is broken by a negligible amount of thermal
noise, entanglement distribution can be reactivated by a vanishing amount of
separable correlations ($g_{\text{ER}}\simeq\tau\simeq0$).

\begin{figure}[ptbh]
\vspace{-2.3cm}
\par
\begin{center}
\includegraphics[width=0.55\textwidth] {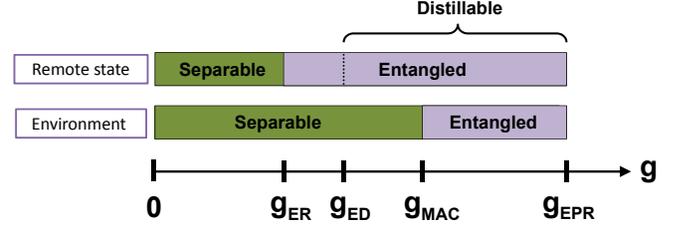}
\end{center}
\par
\vspace{-2.4cm}\caption{Separability properties of the remote state versus
those of the positive AC\ environments ($g>0$). Here we consider the regime of
high transmissivity $\tau\gtrsim0.63$, where remote entanglement can be
distilled even in the presence of separable environments ($g_{\text{ED}}<g\leq
g_{\text{MAC}}$).}%
\label{linea2}%
\end{figure}

Proceeding with our analysis of the AC environments, we see that entanglement
distillation is possible for transmissivities
\begin{equation}
\tau>e^{-1}(\sqrt{e}-1)^{2}\simeq0.15~,
\end{equation}
and values of the correlation parameter $g_{\text{ED}}<g\leq g_{\text{EPR}}$,
where the entanglement-distillation threshold is now
\begin{equation}
g_{\text{ED}}:=\frac{1+\tau-e^{-1}}{1-\tau}\geq g_{\text{ER}}~.
\end{equation}
At high transmissivities $\tau>1-e^{-1}\simeq0.63$, we have $g_{\text{ED}%
}<g_{\text{MAC}}$, which means that distillable entanglement can be restored
in the presence of separable AC\ environments. For a pictorial representation
see Fig.~\ref{linea2}. 

\begin{figure}[ptbh]
\vspace{+0.0cm}
\par
\begin{center}
\includegraphics[width=0.36\textwidth] {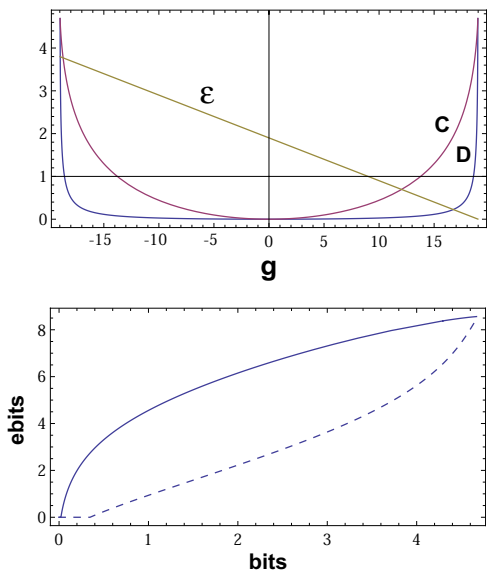}
\end{center}
\par
\vspace{-0.7cm}\caption{\textit{Top panel}. PTS\ eigenvalue $\varepsilon
_{\text{AC}}$ (dimensionless units), Gaussian quantum discord $D$ (dbits) and
purely-classical correlations $C$ (cbits) are plotted versus the correlation
parameter $g$ of the AC environment. Remote entanglement is restored
($\varepsilon_{\text{AC}}<1$)\ only in positive AC environments ($g>0$) and
after critical values of $D$ and $C$. Optimal restoration is achieved for the
positive EPR environment, where $C=D$ are maximal. Numerical example for
$\tau=0.9$ and $\omega=\omega_{EB}=$ $19$. \textit{Bottom panel}. Remote
entanglement quantified by the log-negativity (ebits) is plotted versus
Gaussian discord (dbits, solid line) and classical correlations (cbits, dashed
line) present in the environment. Once entanglement is reactivated, the number
of remote ebits is increasing in the number of environmental dbits and cbits
(here $\tau$ and $\omega$ are chosen as before).}%
\label{discordAC1}%
\end{figure}

Finally, we analyze the behavior of the remote entanglement in terms of the
different types of environmental correlations. As evident from the top panel
of Fig.~\ref{discordAC1}, remote entanglement starts to be restored
($\varepsilon_{\text{AC}}<1$) only after certain values of Gaussian discord
and classical correlations are present in the positive AC\ environment. After
these critical values, the number of remote ebits of log-negativity are
increasing in both correlations, as shown in the bottom panel of
Fig.~\ref{discordAC1}. Curves refer to the same numerical example as before
($\tau=0.9$) and similar behavior holds at the other transmissivities. In
general, the positive AC\ environments outperform the SC\ environments in
restoring entanglement, as it is clear by comparing the curves of
Fig.~\ref{discordAC1} to those of Fig.~\ref{discordSC1}. Such a better
performance can also be recognized from the asymmetric shape of the
reactivation area in Fig.~\ref{dirTOT}. 

\begin{figure}[ptbh]
\vspace{-0.5cm}
\par
\begin{center}
\includegraphics[width=0.45\textwidth] {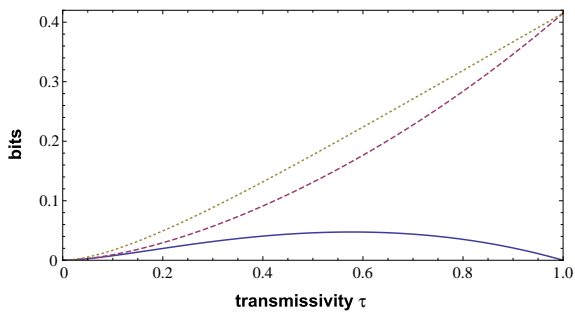}
\end{center}
\par
\vspace{-0.6cm}\caption{For positive AC environments at any $\tau$, we plot
the critical number of correlation bits after which remote entanglement starts
to be restored. We show the critical numbers of dbits (solid), cbits (dashed)
and their total (dotted), corresponding to bits of quantum mutual
information.}%
\label{criticalacPIC}%
\end{figure}

The better performance of these environments is also evident from
Fig.~\ref{criticalacPIC}, where we plot the critical number of correlation
bits which are needed for the restoration of remote entanglement (dbits, cbits
and bits of total correlations). We need less than $2-\log3\simeq0.415~$bits
of total correlations to trigger the reactivation at any transmissivity $\tau
$. In particular, the critical number of dbits is always less than $0.05$. For
$\tau$ close to zero, a negligible amount of correlations is needed. This is
compatible with the fact that, for these environments, the threshold for
entanglement-restoration $g_{\text{ER}}$ goes rapidly to zero for
$\tau\rightarrow0$, so that a small $g>g_{\text{ER}}\simeq0$ is sufficient to
reactivate the entanglement distribution in the high loss regime. In the
opposite regime of extremely low loss ($\tau\rightarrow1$) and high thermal
noise ($\omega_{\text{EB}}\rightarrow+\infty$), the critical number of dbits
falls to zero while the numbers of cbits tend to its maximum.

\subsubsection{General non-monotonicity of entanglement
restoration\label{nonMONO}}

According to the previous sections, the restoration of entanglement is a
threshold process which starts only after a critical amount of environmental
correlations is injected into the travelling modes. After this threshold, the
number of ebits remotely recovered is monotonically increasing in the number
of correlation bits (dbits, cbits or total bits). However, this monotonic
behavior holds true as long as we compare environments within the same class,
i.e., we analyze SC environments separately from AC\ environments. In general,
the monotonicity is lost when we compare two arbitrary points in the
correlation plan. For instance, it is sufficient to compare the
MSC\ environment with the (positive) MAC environment to find that entanglement
restoration is not monotonic in any type of correlations, i.e., classical,
quantum or total.

As we have already discussed in Sec.~\ref{subSECspecific}, at any fixed
thermal noise $\omega>1$, MSC\ and MAC environments have exactly the same
amount of classical correlations $C_{\text{MSC}}=C_{\text{MAC}}$, but
different Gaussian discord $D_{\text{MSC}}>D_{\text{MAC}}$\ and, therefore,
different total correlations $I_{\text{MSC}}>I_{\text{MAC}}$. This is because
MAC\ environments are more entropic than MSC environments. Their difference in
entropy%
\begin{equation}
\delta S:=S_{\text{MSC}}-S_{\text{MAC}}=h(2\omega-1)-2h\left(  \sqrt
{2\omega-1}\right)  <0,
\end{equation}
directly quantifies the difference in the correlations, i.e.,
\begin{equation}
D_{\text{MSC}}-D_{\text{MAC}}=I_{\text{MSC}}-I_{\text{MAC}}=-\delta S>0.
\end{equation}

In particular, all these relations hold true at the threshold for
entanglement-breaking $\omega=\omega_{\text{EB}}$, where the number of ebits
remotely restored is bigger for the positive MAC environment, i.e., the
log-negativities satisfy
\begin{equation}
\mathcal{N}_{\text{MSC}}<\mathcal{N}_{\text{MAC}}~,
\end{equation}
which is equivalent to $\varepsilon_{\text{MSC}}>\varepsilon_{\text{MAC}}$,
where the two eigenvalues $\varepsilon_{\text{MSC}}$ and $\varepsilon
_{\text{MAC}}$ are given in Eqs.~(\ref{EPSmsc}) and~(\ref{EPSmac}),
respectively. Thus, in this specific example, an increase in the environmental
correlations, quantified by the Gaussian discord ($D_{\text{MSC}%
}>D_{\text{MAC}}$) or the total correlations ($I_{\text{MSC}}>I_{\text{MAC}}
$), corresponds to a decrease in the number of ebits remotely recovered
($\mathcal{N}_{\text{MSC}}<\mathcal{N}_{\text{MAC}}$). At the same time, we
also see that this decrease happens for constant classical correlations
($D_{\text{MSC}}=D_{\text{MAC}}$).

Such a phenomenon could have different potential explanations. There could be
a problem with Gaussian discord, which could not be equal to the actual
discord for these environments. If this were true, we would have a different
quantification for both quantum correlations $D$ and classical correlations
$C$. Then, monotonicity could be re-established in terms of these actual
quantities. More drastically, it could be that entanglement restoration might
be monotonically connected with a different and more subtle quantification of
the environmental correlations.

It is also important to note that this analysis does not change if we consider
a different quantification for the remote entanglement. For instance, we could
consider the entanglement of formation~\cite{EoF} in the place of the
log-negativity to quantify the entanglement of Alice and Bob's two-mode state.
However, such a state turns out to be a symmetric Gaussian state, so that its
entanglement of formation is equal to its Gaussian entanglement of
formation~\cite{Giedke}, which is monotonically related to its
log-negativity~\cite{GerryIll}.

\subsubsection{Evolution of the EPR\ correlations\label{subEPRevolution}}

In order to give another point of view to the dynamics of the process, we
describe the evolution of the EPR\ correlations of Charlie's input state when
the two modes are transmitted through the correlated-noise Gaussian
environment, with transmission $\tau$, thermal noise $\omega$ and correlations
$\mathbf{G}$, as depicted in the left panel of Fig.~\ref{twomodeEB} (see
Appendix~\ref{APP_subEPR} for more details on EPR\ correlations). As we will
see below, the entanglement restored at Alice's and Bob's stations is not
necessarily in the form EPR correlations.

We can easily write the input-output Bogoliubov transformations for the
bosonic modes involved in the process%
\begin{align}
\mathbf{\hat{x}}_{A^{\prime}}  &  =\sqrt{\tau}\mathbf{\hat{x}}_{A}%
+\sqrt{1-\tau}\mathbf{\hat{x}}_{E_{1}}\label{bogo1}\\
\mathbf{\hat{x}}_{B^{\prime}}  &  =\sqrt{\tau}\mathbf{\hat{x}}_{B}%
+\sqrt{1-\tau}\mathbf{\hat{x}}_{E_{2}} \label{bogo2}%
\end{align}
where $\mathbf{\hat{x}}=(\hat{q},\hat{p})^{T}$ is a vector of quadratures.
From these equations, we can extract the output EPR\ operators
\begin{equation}
\hat{q}_{-}^{\prime}:=\frac{\hat{q}_{A^{\prime}}-\hat{q}_{B^{\prime}}}%
{\sqrt{2}}~,~\hat{p}_{+}^{\prime}:=\frac{\hat{p}_{A^{\prime}}+\hat
{p}_{B^{\prime}}}{\sqrt{2}}~.
\end{equation}
Now, using the CM of the input EPR state~(\ref{EPRcmMU}) and that of the
environment~(\ref{EVE_cmAPP}), we can compute their variances%
\begin{align}
\boldsymbol{\Lambda}  &  :=\left(
\begin{array}
[c]{cc}%
V(\hat{q}_{-}^{\prime}) & \\
& V(\hat{p}_{+}^{\prime})
\end{array}
\right) \nonumber\\
&  =\tau(\mu-\mu^{\prime})\mathbf{I}+(1-\tau)(\omega\mathbf{I}-\mathbf{ZG})~.
\end{align}
In the limit of $\mu\gg1$, we have%
\begin{equation}
\boldsymbol{\Lambda}\rightarrow\boldsymbol{\Lambda}_{\infty}=(1-\tau
)(\omega\mathbf{I}-\mathbf{ZG})~, \label{EPRheis}%
\end{equation}
and assuming entanglement-breaking ($\omega=\omega_{\text{EB}}$) we get%
\begin{equation}
\boldsymbol{\Lambda}_{\infty,\text{EB}}=(1+\tau)\mathbf{I}-(1-\tau
)\mathbf{ZG}~. \label{EPRheis2}%
\end{equation}
Analyzing Eq.~(\ref{EPRheis2}) we see that the EPR\ condition
$\boldsymbol{\Lambda}_{\infty,\text{EB}}<\mathbf{I}$ can be realized by
suitable choices of the correlation block $\mathbf{G}$ (here the
EPR\ condition corresponds to having quadrature correlations between the
output modes, quantified by the variances $V(\hat{q}_{-}^{\prime})$ and
$V(\hat{p}_{+}^{\prime})$, which are below the unit-variance of the vacuum noise).

Let us explicitly compare the different types of environments. For the
memoryless environment ($\mathbf{G}=\mathbf{0}$) we have $\boldsymbol{\Lambda
}_{\infty,\text{EB}}=(1+\tau)\mathbf{I}$ which means that the EPR\ variances
are always greater than or equal to one, i.e., EPR\ correlations do not survive.

For the AC environment ($\mathbf{G}=g\mathbf{Z}$) we have
\begin{equation}
\boldsymbol{\Lambda}_{\infty,\text{EB}}=[(1+\tau)-(1-\tau)g]\mathbf{I~}.
\end{equation}
It is easy to check that the EPR\ condition $\boldsymbol{\Lambda}%
_{\infty,\text{EB}}<\mathbf{I}$ is achieved by physical values of
$g>g_{\text{ER}}$, where $g_{\text{ER}}$ is given in Eq.~(\ref{gEGasymm}).
This means that the remote entanglement generated by this environment is
always in the form of EPR\ correlations (of the same type of the original EPR
state at Charlie's station).

For the SC environment ($\mathbf{G}=g\mathbf{I}$) we have
\begin{equation}
\boldsymbol{\Lambda}_{\infty,\text{EB}}=(1+\tau)\mathbf{I}-(1-\tau
)g\mathbf{Z~,}%
\end{equation}
and we can check that the condition $\boldsymbol{\Lambda}_{\infty,\text{EB}%
}<\mathbf{I}$ is not realizable by any choice of $g$. Thus, the initial
EPR\ correlations do not survive in this case. Nevertheless we have explicitly
proven that remote entanglement can be distributed in the presence of this environment.

\section{Conclusion and discussion~\label{SECconclusion}}

In conclusion, we have shown how the combination of two entanglement-breaking
channels into a joint correlated-noise environment can restore the
distribution of entanglement. Interestingly, this reactivation is successfully
induced by the presence of separable correlations and occurs for quantum
systems with Hilbert spaces of any dimension (qubits, qudits and bosonic modes).

In our first analysis, we have considered twirling environments which are
classically-correlated, being realizable by random LOCCs and described by
zero-discord environmental states. We have considered correlated Pauli
environments for qubits, multidimensional twirling environments for qudits
and, finally, non-Gaussian bosonic environments based on phase-space
rotations. All these scenarios are characterized by one-system
entanglement-breaking (so that Charlie cannot share any entanglement with
Alice or Bob), and two-system entanglement-preserving (so that Charlie is able
to transmit entanglement to Alice and Bob).

Achieving these simultaneous conditions is somehow easy for these
environments, since there are well-known classes of entangled states which are
invariant under their action (thus forming decoherence-free subspaces).
Depending on the type of twirling, i.e., $U\otimes U$ or $U\otimes U^{\ast}$,
these invariant states are Werner or isotropic states, respectively. In the
case of bosonic systems, we can find entangled states which are invariant
under\ the action of anti-correlated phase-space rotations $R_{\theta}\otimes
R_{-\theta}$, and they coincide with the continuous-variable Werner states.

More interestingly, in our second analysis on bosonic systems in Gaussian
environments, we have shown how entanglement distribution is still possible
even in the absence of decoherence-free subspaces (a similar relaxation of
conditions can also be found in the context of quantum error
correction~\cite{Beny}). In this case, entanglement is no longer preserved but
still can survive the double transmission and be distilled by the remote parties.

More precisely, we have considered a realistic model of Gaussian environment
with correlated noise, which generalizes the standard memoryless environment
with losses and thermal noise. Then, we have imposed the condition of one-mode
entanglement-breaking by choosing a sufficiently high value of thermal noise
$\omega_{\text{EB}}$ for any value of the transmission (so that Charlie can
never share entanglement with the remote parties). Under these conditions of
extreme decoherence, we have shown that a distillable amount of entanglement
is able to survive the two-mode transmission and reach Alice's and Bob's
stations. This reactivation of entanglement distribution is a direct effect of
the correlations which are injected by the Gaussian environment. Remarkably,
the injection of separable correlations is sufficient to break the mechanism
of entanglement-breaking and trigger the reactivation.

In a further analysis, we have shown that remote entanglement starts to be
restored only after a critical amount of correlations is present in the
environment, where these correlations can be quantified in terms of Gaussian
discord, classical correlations and total correlations (quantum mutual
information). After this critical amount has been reached, the number of
remote ebits is not always increasing in the number of correlation bits
injected by the environment, despite such a monotonicity being true when we
restrict the environments to classes with specific types of noise correlations
(SC\ and AC\ environments). It is possible that the monotonicity could be
extended to the entire correlation plan if Gaussian discord were proven to be
strictly larger than the actual discord. More drastically, the monotonicity
could\ hold in terms of a different and unknown quantification of the
environmental correlations.

From a theoretical point of view, the fact that separability can be exploited
to recover from entanglement-breaking is a paradoxical behavior which poses
fundamental questions on the intimate relations between local and nonlocal
correlations. While the reactivation process is easily understandable in the
presence of entangled environments, where the original system entanglement is
partly or fully replaced by the environmental entanglement, its interpretation
is puzzling in the case of separable environments, where no injection of
entanglement may take place.

In conclusion, in terms of potential practical impact, our analysis shows new
perspectives for entanglement distribution and distillation in conditions of
extreme decoherence as long as the presence of noise correlations and memory
effects can be identified in the environment, such as those arising from
non-Markovian dynamics of open quantum systems~\cite{Petruccione,Weiss}.
According to our findings, such correlations and effects do not need to be
strong, since separable correlations and classical memories may be sufficient
to restore broken entanglement.

Note that memory channels and non-Markovian environments are ubiquitous. For
instance, they naturally arise in the context of spin chains~\cite{Bose03} and
micromasers~\cite{Benenti09}. Other important examples can be found in
condensed matter, in particular when we consider the dynamics of quantum dots
in photonic crystals~\cite{Madsen11}. In the bosonic setting, memory Gaussian
channels come into play when electromagnetic modes propagate through
dispersive media, such as linear optical systems or free-space. In this case,
correlations and memory effects are naturally introduced by
diffraction~\cite{Shapiro09,Lupo1,Lupo2,qreadDIFF}. Finally, other examples of
bosonic memory channels can also be found in the propagation of
electromagnetic radiation through atmospheric
turbulence~\cite{Tyler09,Semenov09,Boyd11}.

\section{Acknowledgements}

This work has been supported by EPSRC under the research grant HIPERCOM
(EP/J00796X/1) and by The Leverhulme Trust. Special thanks of the author are
for S. L. Braunstein, M. Paternostro and C. Ottaviani. The author would also
like to thank (in random order) P. Horodecki, O. Oreshkov, A. Furusawa, G.
Spedalieri, G. Adesso, M. Gu, S. Mancini, G. Chiribella, O. Hirota, B. Munro,
N. Metwally, R. Namiki, P. Tombesi, S. Guha, M. J. W. Hall, S. Danilishin, M.
Barbieri, M. Bellini, R. Filip, and J. Eschner for discussions and comments.

\appendix

\section{Miscellaneous proofs\label{APPmisce}}
Here we report some of the proofs related to our analysis of
entanglement preservation in twirling environments. This appendix
contains new results but also known facts which are given to the
reader for the sake of completeness (e.g., Sec.~\ref{DephaseAPP}).

\subsection{Entanglement-breaking conditions for qubit depolarizing channels
\label{DEPapp}}

Here we show the conditions under which the depolarizing channel of
Eq.~(\ref{qubitDEPO}) becomes an entanglement-breaking channel. This means to
find a specific regime for the probabilities $\{p_{k}\}_{k=0}^{3}$
characterizing the channel.

First of all note that, for Hilbert spaces of finite dimension $d$, a simple
way to check if a quantum channel $\mathcal{E}$ is entanglement-breaking is to
test it on the maximally-entangled state $\left\vert \psi\right\rangle _{AB}$
of Eq.~(\ref{maxENTstate}). In other words, if $(\mathcal{E}_{A}%
\otimes\mathcal{I}_{B})\left\vert \psi\right\rangle _{AB}\left\langle
\psi\right\vert $ is separable, then $(\mathcal{E}_{A}\otimes\mathcal{I}%
_{B})\rho_{AB}$ is separable for any input state $\rho_{AB}$~\cite{EBchannels}%
. In the case of qubits, we can test the channel on the triplet state
$\left\vert +\right\rangle _{AB}$ of Eq.~(\ref{tripletSTATE}). We then compute
the output state%
\begin{align}
\Phi &  :=(\mathcal{E}_{A}\otimes\mathcal{I}_{B})(\left\vert +\right\rangle
_{AB}\left\langle +\right\vert )\nonumber\\
&  =\sum_{k=0}^{3}p_{k}(P_{k}\otimes I)\left\vert +\right\rangle
_{AB}\left\langle +\right\vert (P_{k}^{\dagger}\otimes I)\nonumber\\
&  =p_{0}\left\vert +\right\rangle _{AB}\left\langle +\right\vert \nonumber\\
&  +p_{1}(X\otimes I)\left\vert +\right\rangle _{AB}\left\langle +\right\vert
(X\otimes I)\\
&  +p_{2}(Y\otimes I)\left\vert +\right\rangle _{AB}\left\langle +\right\vert
(Y^{\dagger}\otimes I)\nonumber\\
&  +p_{3}(Z\otimes I)\left\vert +\right\rangle _{AB}\left\langle +\right\vert
(Z\otimes I)~.
\end{align}
Adopting the computational basis $\{\left\vert 00\right\rangle ,\left\vert
01\right\rangle ,\left\vert 10\right\rangle ,\left\vert 11\right\rangle \}$
and using $X\left\vert u\right\rangle =\left\vert u\oplus1\right\rangle $,
$Z\left\vert u\right\rangle =(-1)^{u}\left\vert u\right\rangle $ and $Y=iXZ$,
we get%
\begin{equation}
\Phi=\sum_{ijkl}\Phi_{ijkl}\left\vert i,j\right\rangle _{AB}\left\langle
k,l\right\vert ~,
\end{equation}
where the coefficients $\Phi_{ijkl}=\left\langle i,j\right\vert \Phi\left\vert
k,l\right\rangle $ are the elements of the following density matrix%
\begin{equation}
\boldsymbol{\Phi}=\frac{1}{2}\left(
\begin{array}
[c]{cccc}%
p_{0}+p_{3} & 0 & 0 & p_{0}-p_{3}\\
0 & p_{1}+p_{2} & p_{1}-p_{2} & 0\\
0 & p_{1}-p_{2} & p_{1}+p_{2} & 0\\
p_{0}-p_{3} & 0 & 0 & p_{0}+p_{3}%
\end{array}
\right)  ~.
\end{equation}
To check the separability properties we adopt the Peres-Horodecki
criterion~\cite{PERES,HOROsep}. This corresponds to compute the partial
transpose (PT) of the state which is given by the following linear map
\begin{align}
\Phi &  =\sum_{ijkl}\Phi_{ijkl}\left\vert i\right\rangle _{A}\left\langle
k\right\vert \otimes\left\vert j\right\rangle _{B}\left\langle l\right\vert
\nonumber\\
&  \rightarrow\mathrm{PT}(\Phi)=\sum_{ijkl}\Phi_{ijkl}\left\vert
i\right\rangle _{A}\left\langle k\right\vert \otimes(\left\vert j\right\rangle
_{B}\left\langle l\right\vert )^{T}\nonumber\\
&  =\sum_{ijkl}\Phi_{ijkl}\left\vert i\right\rangle _{A}\left\langle
k\right\vert \otimes\left\vert l\right\rangle _{B}\left\langle j\right\vert
\nonumber\\
&  =\sum_{ijkl}\Phi_{ilkj}\left\vert i\right\rangle _{A}\left\langle
k\right\vert \otimes\left\vert j\right\rangle _{B}\left\langle l\right\vert ~.
\end{align}
At the level of the density matrix we then have
\begin{equation}
\boldsymbol{\Phi}=((\Phi_{ijkl}))\rightarrow\mathrm{PT}(\boldsymbol{\Phi
})=((\Phi_{ilkj}))~,
\end{equation}
i.e., $j\longleftrightarrow l$ swapping. It is easy to check that the
partially-transposed matrix
\begin{equation}
\mathrm{PT}(\boldsymbol{\Phi})=\frac{1}{2}\left(
\begin{array}
[c]{cccc}%
p_{0}+p_{3} & 0 & 0 & p_{1}-p_{2}\\
0 & p_{1}+p_{2} & p_{0}-p_{3} & 0\\
0 & p_{0}-p_{3} & p_{1}+p_{2} & 0\\
p_{1}-p_{2} & 0 & 0 & p_{0}+p_{3}%
\end{array}
\right)
\end{equation}
has eigenvalues
\begin{equation}
\lambda_{k}=\frac{1}{2}-p_{k}~~(k=0,1,2,3)~.
\end{equation}
Thus the partially-transposed state has the following spectral decomposition%
\begin{equation}
\mathrm{PT}(\Phi)=\sum_{k=0}^{3}\left(  \frac{1}{2}-p_{k}\right)  \left\vert
\eta_{k}\right\rangle \left\langle \eta_{k}\right\vert ~,
\end{equation}
with $\left\vert \eta_{k}\right\rangle $ orthogonal eigenstates. This operator
is positive ($\geq0$) if and only if%
\begin{equation}
p_{k}\leq\frac{1}{2}~,~\text{for}~k=0,1,2,3~.
\end{equation}
As a result, the output state $\Phi$ is separable (and the channel is
entanglement-breaking) if and only if $p_{k}\leq1/2$ for every $k$, as
reported in the main text.

\subsection{Unitary $2$-design for the $U\otimes U^{\ast}$ twirling channel
\label{TwirlAVEapp}}

Let us consider two qudits with the same dimension $d$, so that the composite
system is described by an Hilbert space $\mathcal{H}=\mathcal{H}_{A}%
\otimes\mathcal{H}_{B}$ with finite dimension $d^{2}$. From the
literature~\cite{Designs,Gross}, we know that we can write the following
equality for any input state $\rho_{AB}$ subject to a $U\otimes U$ twirling
channel%
\begin{align}
\mathcal{E}_{UU}(\rho_{AB})  &  :=\int_{\mathcal{U}(d)}dU~(U\otimes
U)\rho_{AB}(U\otimes U)^{\dagger}\label{Euu1}\\
&  =\frac{1}{K}\sum_{k=0}^{K-1}(U_{k}\otimes U_{k})\rho_{AB}(U_{k}\otimes
U_{k})^{\dagger}~, \label{Euu2}%
\end{align}
which is valid for $U_{k}\in\mathcal{D}$, where $\mathcal{D}$\ is a unitary
2-design with $K$ elements. Here we can easily show that
\begin{align}
\mathcal{E}_{UU^{\ast}}(\rho_{AB})  &  :=\int_{\mathcal{U}(d)}dU~(U\otimes
U^{\ast})\rho_{AB}(U\otimes U^{\ast})^{\dagger}\label{Euustar1}\\
&  =\frac{1}{K}\sum_{k=0}^{K-1}(U_{k}\otimes U_{k}^{\ast})\rho_{AB}%
(U_{k}\otimes U_{k}^{\ast})^{\dagger}~, \label{Euustar2}%
\end{align}
where $U_{k}$ belongs to the same design $\mathcal{D}$ as before. In a few
words, the two twirling channels, $\mathcal{E}_{UU}$ and $\mathcal{E}%
_{UU^{\ast}}$, can be decomposed using the same unitary 2-design.

In the first step of the proof we show that $\mathcal{E}_{UU}$ and
$\mathcal{E}_{UU^{\ast}}$ are connected by a partial transposition. Consider
an arbitrary input state $\rho_{AB}$ decomposed in the orthonormal basis of
$\mathcal{H}$~\cite{Notation}%
\begin{equation}
\rho_{AB}=\sum_{ijkl}\rho_{ij}^{kl}\left\vert i\right\rangle _{A}\left\langle
j\right\vert \otimes\left\vert k\right\rangle _{B}\left\langle l\right\vert ~.
\end{equation}
Its partial transpose corresponds to transposing system $B$ only, i.e.,%
\begin{equation}
\mathrm{PT}(\rho_{AB})=\sum_{ijkl}\rho_{ij}^{kl}\left\vert i\right\rangle
_{A}\left\langle j\right\vert \otimes\left\vert l\right\rangle _{B}%
\left\langle k\right\vert ~.
\end{equation}
We first prove that
\begin{equation}
\mathcal{E}_{UU^{\ast}}(\rho_{AB})=\mathrm{PT}\{\mathcal{E}_{UU}%
[\mathrm{PT}(\rho_{AB})]\}~. \label{PTconnection}%
\end{equation}
In fact, by linearity we have
\begin{gather}
\mathcal{E}_{UU}[\mathrm{PT}(\rho_{AB})]=\int_{\mathcal{U}(d)}dU~(U\otimes
U)\mathrm{PT}(\rho_{AB})(U\otimes U)^{\dagger}\nonumber\\
=\sum_{ijkl}\rho_{ij}^{kl}\int_{\mathcal{U}(d)}dU~U\left\vert i\right\rangle
_{A}\left\langle j\right\vert U^{\dagger}\otimes U\left\vert l\right\rangle
_{B}\left\langle k\right\vert U^{\dagger}.
\end{gather}
Since $[U\left\vert l\right\rangle _{B}\left\langle k\right\vert U^{\dagger
}]^{T}=U^{\ast}\left\vert k\right\rangle _{B}\left\langle l\right\vert
(U^{\ast})^{\dagger}$ we get%
\begin{gather}
\mathrm{PT}\{\mathcal{E}_{UU}[\mathrm{PT}(\rho_{AB})]\}=\nonumber\\
=\sum_{ijkl}\rho_{ij}^{kl}\int_{\mathcal{U}(d)}dU~U\left\vert i\right\rangle
_{A}\left\langle j\right\vert U^{\dagger}\otimes U^{\ast}\left\vert
k\right\rangle _{B}\left\langle l\right\vert (U^{\ast})^{\dagger}\nonumber\\
=\int_{\mathcal{U}(d)}dU(U\otimes U^{\ast})\rho_{AB}(U\otimes U^{\ast
})^{\dagger}=\mathcal{E}_{UU^{\ast}}(\rho_{AB})~.
\end{gather}

Now the second step is to combine Eq.~(\ref{PTconnection}) with the unitary
$2$-design for $\mathcal{E}_{UU}$. First it is important to note that the
equivalence between Eqs.~(\ref{Euu1}) and~(\ref{Euu2}) is valid not only when
$\rho_{AB}$ is a density operator but, more generally, when it is an Hermitian
linear operator. This extension is straightforward to prove. Suppose that the
linear operator $O:\mathcal{H}\rightarrow\mathcal{H}$ is Hermitian. Then its
spectral decomposition involves real eigenvalues and orthonormal eigenvectors,
i.e., we can write%
\begin{equation}
O=\sum_{n}O_{n}\left\vert \phi_{n}\right\rangle \left\langle \phi
_{n}\right\vert ~,
\end{equation}
where $O_{n}\in\mathbb{R}$ and $\left\langle \phi_{n}\right\vert \left.
\phi_{m}\right\rangle =\delta_{nm}$. Now we can write
\begin{gather}
\mathcal{E}_{UU}(O):=\int_{\mathcal{U}(d)}dU~(U\otimes U)O(U\otimes
U)^{\dagger}\label{s1}\\
=\sum_{n}O_{n}\int_{\mathcal{U}(d)}dU~(U\otimes U)\left\vert \phi
_{n}\right\rangle \left\langle \phi_{n}\right\vert (U\otimes U)^{\dagger
}\label{s2}\\
=\sum_{n}O_{n}\frac{1}{K}\sum_{k=0}^{K-1}(U_{k}\otimes U_{k})\left\vert
\phi_{n}\right\rangle \left\langle \phi_{n}\right\vert (U_{k}\otimes
U_{k})^{\dagger}\label{s3}\\
=\frac{1}{K}\sum_{k=0}^{K-1}(U_{k}\otimes U_{k})O(U_{k}\otimes U_{k}%
)^{\dagger}~, \label{s4}%
\end{gather}
where (\ref{s1})$\rightarrow$(\ref{s2}) by linearity, (\ref{s2})$\rightarrow
$(\ref{s3}) by the fact that $\left\vert \phi_{n}\right\rangle \left\langle
\phi_{n}\right\vert $ are projectors (and therefore states) and (\ref{s3}%
)$\rightarrow$(\ref{s4}) by linearity again.

As a result, we can apply the equivalence between Eqs.~(\ref{s1})
and~(\ref{s4})\ to the linear operator $\mathrm{PT}(\rho_{AB})$ which fails to
be a density operator when $\rho_{AB}$ is entangled but still it is Hermitian
(and unit trace) in the general case. Thus, we can write%
\begin{gather}
\mathcal{E}_{UU}[\mathrm{PT}(\rho_{AB})]=\frac{1}{K}\sum_{t=0}^{K-1}%
(U_{t}\otimes U_{t})\mathrm{PT}(\rho_{AB})(U_{t}\otimes U_{t})^{\dagger
}\nonumber\\
=\frac{1}{K}\sum_{t=0}^{K-1}\sum_{ijkl}\rho_{ij}^{kl}U_{t}\left\vert
i\right\rangle _{A}\left\langle j\right\vert U_{t}^{\dagger}\otimes
U_{t}\left\vert l\right\rangle _{B}\left\langle k\right\vert U_{t}^{\dagger}~.
\end{gather}
Now, using the connection in Eq.~(\ref{PTconnection}) and the fact that
$[U_{t}\left\vert l\right\rangle _{B}\left\langle k\right\vert U_{t}^{\dagger
}]^{T}=U_{t}^{\ast}\left\vert k\right\rangle _{B}\left\langle l\right\vert
(U_{t}^{\ast})^{\dagger}$, we can write%
\begin{gather}
\mathcal{E}_{UU^{\ast}}(\rho_{AB})=\mathrm{PT}\{\mathcal{E}_{UU}%
[\mathrm{PT}(\rho_{AB})]\}\nonumber\\
=\frac{1}{K}\sum_{t=0}^{K-1}\sum_{ijkl}\rho_{ij}^{kl}U_{t}\left\vert
i\right\rangle _{A}\left\langle j\right\vert U_{t}^{\dagger}\otimes
U_{t}^{\ast}\left\vert k\right\rangle _{B}\left\langle l\right\vert
(U_{t}^{\ast})^{\dagger}\nonumber\\
=\frac{1}{K}\sum_{t=0}^{K-1}(U_{t}\otimes U_{t}^{\ast})\rho_{AB}(U_{t}\otimes
U_{t}^{\ast})^{\dagger}~, \label{finaPT}%
\end{gather}
which gives the equivalence between Eqs.~(\ref{Euustar1}) and~(\ref{Euustar2}).

\subsection{Partial Haar average of a linear operator\label{HaarAVEsec}}

In this short appendix we give a simple proof of Eq.~(\ref{depola2}). Consider
an Hilbert space $\mathcal{H}=\mathcal{H}_{A}\otimes\mathcal{H}_{B}$ with
finite dimension $d_{A}d_{B}$ (where $d_{A}$ and $d_{B}$ are generally
different). Given a linear operator $T:\mathcal{H}\rightarrow\mathcal{H}$, we
can always decompose it in an orthonormal basis
\begin{equation}
T=\sum_{ijkl}T_{ij}^{kl}\left\vert i\right\rangle _{A}\left\langle
j\right\vert \otimes\left\vert k\right\rangle _{B}\left\langle l\right\vert ~.
\end{equation}
Then, we can write the following partial Haar average, where only system $A$
is averaged on the unitary group%
\begin{align}
\left\langle T\right\rangle _{U\otimes I}  &  :=\int_{\mathcal{U}%
(d)}dU(U\otimes I)T(U^{\dagger}\otimes I)\nonumber\\
&  =\sum_{ijkl}T_{ij}^{kl}\left(  \int_{\mathcal{U}(d)}dU~U\left\vert
i\right\rangle _{A}\left\langle j\right\vert U^{\dagger}\right)
\otimes\left\vert k\right\rangle _{B}\left\langle l\right\vert . \label{TUI}%
\end{align}
Now we use Eq.~(\ref{twirlingID}) with linear operator $O=\left\vert
i\right\rangle _{A}\left\langle j\right\vert $, which gives%
\begin{equation}
\int_{\mathcal{U}(d)}dU~U\left\vert i\right\rangle _{A}\left\langle
j\right\vert U^{\dagger}=\delta_{ij}\frac{I}{d_{A}}~.
\end{equation}
Then, by replacing this expression in Eq.~(\ref{TUI}), we get%
\begin{align}
\left\langle T\right\rangle _{U\otimes I}  &  =\frac{I}{d_{A}}\otimes
\sum_{ikl}T_{ii}^{kl}\left\vert k\right\rangle _{B}\left\langle l\right\vert
\nonumber\\
&  =\frac{I}{d_{A}}\otimes\mathrm{Tr}_{A}\left(  T\right)  ~. \label{HaarAVE2}%
\end{align}
This is a simple extension of Eq.~(\ref{twirlingID}) to considering the
presence of a second (unaveraged) system $B$. In particular, for $T$ density
operator, we have the result of Eq.~(\ref{depola2}).

\subsection{Uniformly dephasing channel is
entanglement-breaking\label{DephaseAPP}}

Here we prove that the uniformly dephasing channel of Eq.~(\ref{UNIdepha}) is
entanglement-breaking. This is a simple proof of a very intuitive fact, which
is given to the reader for completeness. First of all, let us consider a pure
input state $\rho_{AB}=\left\vert \varphi\right\rangle _{AB}\left\langle
\varphi\right\vert $, expressed in the Fock basis of the two modes as
\begin{equation}
\left\vert \varphi\right\rangle _{AB}=\sum_{kj}c_{kj}\left\vert k\right\rangle
_{A}\otimes\left\vert j\right\rangle _{B}~,~\sum_{kj}\left\vert c_{kj}%
\right\vert ^{2}=1~.
\end{equation}
Since $R_{\theta}\left\vert k\right\rangle =\exp(-i\theta k)\left\vert
k\right\rangle $, we get%
\begin{align}
\rho_{A^{\prime}B}  &  =\sum_{kjk^{\prime}j^{\prime}}c_{kj}c_{k^{\prime
}j^{\prime}}^{\ast}\left(  \int\frac{d\theta}{2\pi}e^{-i\theta(k-k^{\prime}%
)}\left\vert k\right\rangle _{A}\left\langle k^{\prime}\right\vert \right)
\otimes\left\vert j\right\rangle _{B}\left\langle j^{\prime}\right\vert
\nonumber\\
&  =\sum_{kjk^{\prime}j^{\prime}}c_{kj}c_{k^{\prime}j^{\prime}}^{\ast}%
\delta(k-k^{\prime})\left\vert k\right\rangle _{A}\left\langle k\right\vert
\otimes\left\vert j\right\rangle _{B}\left\langle j^{\prime}\right\vert
\nonumber\\
&  =\sum_{kjj^{\prime}}c_{kj}c_{kj^{\prime}}^{\ast}\left\vert k\right\rangle
_{A}\left\langle k\right\vert \otimes\left\vert j\right\rangle _{B}%
\left\langle j^{\prime}\right\vert ~.
\end{align}
Then, by re-distributing the sum, we get%
\begin{align}
\rho_{A^{\prime}B}  &  =\sum_{k}\left(  \sum_{j}c_{kj}\left\vert
kj\right\rangle \right)  \otimes\left(  \sum_{j^{\prime}}c_{kj^{\prime}}%
^{\ast}\left\langle kj^{\prime}\right\vert \right) \nonumber\\
&  =\sum_{k}d_{k}\left(  \sum_{j}\frac{c_{kj}}{\sqrt{d_{k}}}\left\vert
kj\right\rangle \right)  \otimes\left(  \sum_{j^{\prime}}\frac{c_{kj^{\prime}%
}^{\ast}}{\sqrt{d_{k}}}\left\langle kj^{\prime}\right\vert \right)  ~,
\end{align}
where we have introduced $d_{k}:=\sum_{j}\left\vert c_{kj}\right\vert ^{2}$ in
the last step (clearly $\sum_{k}d_{k}=1$). Now introducing the pure state%
\begin{equation}
\left\vert \eta_{k}\right\rangle :=\sum_{j}\frac{c_{kj}}{\sqrt{d_{k}}%
}\left\vert kj\right\rangle ,
\end{equation}
we can write the following spectral decomposition for the output state%
\begin{equation}
\rho_{A^{\prime}B}=\sum_{k}d_{k}\left\vert \eta_{k}\right\rangle \left\langle
\eta_{k}\right\vert ~.
\end{equation}
In particular, note that we can always write the tensor product
\begin{equation}
\left\vert \eta_{k}\right\rangle =\left\vert k\right\rangle \otimes\left\vert
\xi(k)\right\rangle ,~\left\vert \xi(k)\right\rangle :=\sum_{j}\frac{c_{kj}%
}{\sqrt{d_{k}}}\left\vert j\right\rangle ,
\end{equation}
so that the output state is manifestly in separable form
\begin{equation}
\rho_{A^{\prime}B}=\sum_{k}d_{k}\left\vert k\right\rangle _{A^{\prime}%
}\left\langle k\right\vert \otimes\left\vert \xi(k)\right\rangle
_{B}\left\langle \xi(k)\right\vert ~.
\end{equation}
Proof can trivially be extended to mixed states via their spectral
decomposition into pure states.

\subsection{$R_{\theta}\otimes R_{\theta}$-invariant Gaussian states are
separable\label{invGAUSapp}}

To derive the CM of Eq.~(\ref{CM_appfinite}) just check that a $2\times2$ real
matrix $\mathbf{M}$ is invariant under rotations $\mathbf{R}_{\theta
}\mathbf{MR}_{\theta}^{T}=\mathbf{M}$ (or, equivalently, it commutes with
rotations $[\mathbf{M},\mathbf{R}_{\theta}]=0$) if and only if it takes the
asymmetric form%
\begin{equation}
\mathbf{M}=\left(
\begin{array}
[c]{cc}%
m & t\\
-t & m
\end{array}
\right)  ~,
\end{equation}
with $m,t$ real numbers. Thus, the $\mathbf{A},\mathbf{B},\mathbf{C}$ blocks
of the CM~(\ref{CM_appfinite}) must have this general form, with the blocks
$\mathbf{A}$ and $\mathbf{B}$ diagonal by imposing the additional condition of symmetry.

Then, it is easy to check that CM of Eq.~(\ref{CM_appfinite}) describes a
separable Gaussian state. In fact, using suitable local rotations
$\mathbf{R}_{x}\mathbf{\oplus R}_{y}$ (therefore not changing the separability
properties of the state), we can transform $\mathbf{V}_{AB}$ into the simpler
form%
\begin{equation}
\mathbf{V}_{AB}^{\prime}(\alpha,\beta,\gamma)=\left(
\begin{array}
[c]{cc}%
\alpha\mathbf{I} & \gamma\mathbf{I}\\
\gamma\mathbf{I} & \beta\mathbf{I}%
\end{array}
\right)  ~, \label{VabprimeAPP}%
\end{equation}
where $\gamma=\sqrt{\omega^{2}+\varphi^{2}}$. Without loss of generality,
suppose that $\beta\geq\alpha$ and set $\beta-\alpha:=\delta$.

We can always generate $\rho_{AB}^{\prime}$ with CM\ $\mathbf{V}_{AB}^{\prime
}(\alpha,\beta,\gamma)$ by applying local Gaussian channels $\mathcal{I}%
_{A}\otimes\mathcal{G}_{B}$ to the symmetric Gaussian state $\rho_{AB}%
^{\prime\prime}$ with CM $\mathbf{V}_{AB}^{\prime\prime}=\mathbf{V}%
_{AB}^{\prime}(\alpha,\alpha,\gamma)$. It is sufficient to choose the identity
channel $\mathcal{I}_{A}$ and a Gaussian channel $\mathcal{G}_{B}$ with
additive noise $\delta$ (also known as canonical B2 form~\cite{RMP}). It is
now trivial to check that the state $\rho_{AB}^{\prime\prime}$ is separable.
In fact, $\mathbf{V}_{AB}^{\prime\prime}$ is a bona-fide quantum CM when its
parameters $\alpha$ and $\gamma$ satisfy the conditions $\alpha\geq1$ and
$|\gamma|\leq\alpha-1$. Then, we can check that the partially-transposed
symplectic eigenvalues of $\mathbf{V}_{AB}^{\prime\prime}$ are greater than
$1$, i.e., the state is separable, when $|\gamma|\leq\sqrt{\alpha^{2}-1}$,
which is a condition always satisfied. Finally, since $\rho_{AB}^{\prime
\prime}$ is separable, then also $\rho_{AB}^{\prime}$ and $\rho_{AB}$ must be
separable (local operations cannot create entanglement).

\subsection{No entangled Gaussian state is invariant under $U\otimes U$
Gaussian twirlings\label{APPgtwirl}}

The proof is easy. Suppose that we have an arbitrary Gaussian state $\rho
_{AB}$, with mean value $\mathbf{\bar{x}}_{AB}$ and covariance CM
$\mathbf{V}_{AB}$. The action of two Gaussian unitaries $U\otimes U$ on
$\rho_{AB}$ corresponds to apply two identical displacements to its mean
value
\begin{equation}
\mathbf{\bar{x}}_{AB}\rightarrow\mathbf{\bar{x}}_{AB}+(\mathbf{d}%
,\mathbf{d})^{T}~,
\end{equation}
and two identical symplectic matrices to its CM
\begin{equation}
\mathbf{V}_{AB}\rightarrow(\mathbf{S\oplus S})\mathbf{V}_{AB}(\mathbf{S\oplus
S})^{T}~.
\end{equation}
In general, there is no chance to find an invariant Gaussian state, since any
nonzero displacement $\mathbf{d}$ maps the input state into a different output state.

We then restrict the search to considering canonical Gaussian unitaries
($\mathbf{d=0}$) which are one-to-one with the symplectic transformations.
According to Euler's decomposition~\cite{RMP}, any single-mode symplectic
transformation $\mathbf{S}$ can be decomposed into an orthogonal rotation
$\mathbf{R}_{\theta}$, a single-mode squeezing $\mathbf{S}_{r}$, and another
orthogonal rotation $\mathbf{R}_{\omega}$, where $\theta,\omega$ are angles
and $r\geq1$ is a squeezing parameter. As long as some squeezing is present
($r>1$), the trace of the CM changes (physically this corresponds to increase
the mean number of photons). As a result, the CM and, therefore, the Gaussian
state must change.

Thus, we are left to find Gaussian states which are invariant under rotations
only (which are passive transformations, i.e., preserving the trace of the
CM). However, we have already seen that, despite they exist, these Gaussian
states must be separable.

\section{Basic notions on bosonic systems and Gaussian
states\label{APPbosonic}}

In this appendix, we provide notions on bosonic systems, Gaussian states and
their properties, which are useful for understanding the mathematical details
of the main part of the paper. In detail, we discuss Gaussian states and their
statistical moments (Sec.~\ref{APP_subGAUSS}), the symplectic manipulation of
covariance matrices (Sec.~\ref{APP_subSPECTRUM}), the bona-fide conditions for
bipartite Gaussian states and their separability properties
(Sec.~\ref{APP_sub2modes}), EPR correlations (Sec.~\ref{APP_subEPR}) and,
finally, the notions of discord and Gaussian discord
(Sec.~\ref{APP_subDISCORD}).

\subsection{Gaussian states and operations: Representation in terms
of the statistical moments\label{APP_subGAUSS}}

The most important systems in continuous variable quantum information are the
bosonic modes of the electromagnetic field. A bosonic mode is a quantum system
with an infinite-dimensional Hilbert space and described by a pair of
quadrature operators: Position $\hat{q}$ and momentum $\hat{p}$, satisfying
the commutation relation $[\hat{q},\hat{p}]=2i$ (here we set $\hbar=2$). More
generally, a bosonic system of $n$ modes is described by a vector of $2n $
quadrature operators%
\begin{equation}
\mathbf{\hat{x}}^{T}:=(\hat{q}_{1},\hat{p}_{1},\ldots,\hat{q}_{n},\hat{p}%
_{n})~,
\end{equation}
satisfying $[\hat{x}_{i},\hat{x}_{j}]=2i\Omega_{ij}$, where $i,j=1,\ldots,2n$
and $\Omega_{ij}$ is the generic element of the symplectic form%
\begin{equation}
\mathbf{\Omega}:=\bigoplus\limits_{k=1}^{n}\left(
\begin{array}
[c]{cc}%
0 & 1\\
-1 & 0
\end{array}
\right)  ~. \label{Symplectic_Form}%
\end{equation}

In experimental quantum optics, bosonic modes are typically prepared in
Gaussian states. By definition, a quantum state $\rho$ is \textquotedblleft
Gaussian\textquotedblright\ when its Wigner phase-space representation is
Gaussian~\cite{RMP}. A Gaussian state is very easy to describe, being
completely characterized by its first and second-order statistical moments.

The first-order moment is known as the mean value, and defined by
$\mathbf{\bar{x}}:=\langle\mathbf{\hat{x}}\rangle$, where $\langle\hat
{O}\rangle:=\mathrm{Tr}(\hat{O}\rho)$ denotes the average of the arbitrary
operator $\hat{O}$ on the state $\rho$. The second-order moment is known as
the covariance matrix (CM) $\mathbf{V}$, with generic element%
\begin{equation}
V_{ij}:=\frac{1}{2}\left\langle \{\Delta\hat{x}_{i},\Delta\hat{x}%
_{j}\}\right\rangle ~,
\end{equation}
where $\Delta\hat{x}_{i}:=\hat{x}_{i}-\bar{x}_{i}$ is the deviation and
$\{,\}$ is the anticommutator. Note that the diagonal elements $V_{ii}$\ are
just the variances of the quadratures
\begin{equation}
V(\hat{x}_{i}):=\left\langle \Delta\hat{x}_{i}^{2}\right\rangle =\left\langle
\hat{x}_{i}^{2}\right\rangle -\bar{x}_{i}^{2}~.
\end{equation}
The CM\ is a $2n\times2n$ real symmetric matrix, which must satisfy the
uncertainty principle~\cite{SIMONprinc}%
\begin{equation}
\mathbf{V}+i\mathbf{\Omega}\geq0~, \label{unc_PRINC}%
\end{equation}
implying the positive-definiteness $\mathbf{V}>0$~\cite{TwomodePRA}.

The simplest Gaussian states are thermal states. A single-mode thermal state
has zero mean and CM$\ \mathbf{V}=(2\bar{n}+1)\mathbf{I}$, where $\mathbf{I}$
is the $2\times2$ identity matrix and $\bar{n}\geq0$ is the mean number of
thermal photons (vacuum state for $\bar{n}=0$). Multimode thermal states are
constructed by tensor product. Tensor product of states $\rho_{1}\otimes
\rho_{2}$\ corresponds to direct sum of CMs $\mathbf{V}_{1}\oplus
\mathbf{V}_{2}$. Conversely, the partial trace $\rho_{1}=\mathrm{Tr}_{2}%
(\rho_{12})$ corresponds to collapsing the total CM $\mathbf{V}_{12}$ into the
block $\mathbf{V}_{1}$ spanned by $(\hat{q}_{1},\hat{p}_{1})$.

In the experimental practice, Gaussian states are typically processed by
Gaussian operations. The simplest ones are Gaussian unitaries, defined as
those unitary operators which transform Gaussian states into Gaussian states.
At the level of the statistical moments, the action of a Gaussian unitary
$\rho\rightarrow U\rho U^{\dagger}$ corresponds to%
\begin{equation}
\mathbf{\bar{x}}\rightarrow\mathbf{S\bar{x}}+\mathbf{d},~\mathbf{V}%
\rightarrow\mathbf{SVS}^{T},
\end{equation}
where $\mathbf{d}$ is a real displacement vector and $\mathbf{S}$ is a
symplectic matrix, i.e., a real matrix preserving the symplectic form
$\mathbf{S\Omega S}^{T}=\mathbf{\Omega}$. In the Heisenberg picture, a
Gaussian unitary corresponds to the affine map
\begin{equation}
\mathbf{\hat{x}}\rightarrow\mathbf{S\hat{x}}+\mathbf{d~.}%
\end{equation}

In particular, a Gaussian unitary is called \textquotedblleft
canonical\textquotedblright\ when $\mathbf{d}=\mathbf{0}$. The most important
example of canonical unitary is the beam-splitter transformation, which
involves two bosonic modes. This unitary is characterized by the symplectic
matrix%
\begin{equation}
\mathbf{S}(\tau)=\left(
\begin{array}
[c]{cc}%
\sqrt{\tau}\mathbf{I} & \sqrt{1-\tau}\mathbf{I}\\
-\sqrt{1-\tau}\mathbf{I} & \sqrt{\tau}\mathbf{I}%
\end{array}
\right)  ~, \label{BSmatrix}%
\end{equation}
with transmissivity parameter $0\leq\tau\leq1$.

Other important examples of Gaussian operations are the Gaussian
channels~\cite{GaussCH}. These channels describe all those cases where bosonic
systems and environment interact by means of linear and/or bilinear
Hamiltonians. The action of a Gaussian channel $\rho\rightarrow\mathcal{E}%
(\rho)$ corresponds to the following transformation for the CM of the state%
\begin{equation}
\mathbf{V}\rightarrow\mathbf{KVK}^{T}+\mathbf{N}~, \label{TransformationCH}%
\end{equation}
where $\mathbf{K}$ and $\mathbf{N}=\mathbf{N}^{T}$ are $2n\times2n$ real
matrices, satisfying the condition $\mathbf{N}+i\mathbf{\Omega}%
-i\mathbf{K\Omega K}^{T}\geq0$~\cite{Holevo2001}.

The most important example of Gaussian channel is the lossy channel, which is
typically used to model the optical propagation through dissipative linear
media. When a single bosonic mode propagates through a lossy channel, its
CM\ is transformed by Eq.~(\ref{TransformationCH}) with%
\begin{equation}
\mathbf{K}=\sqrt{\tau}\mathbf{I,~N}=(1-\tau)(2\bar{n}+1)\mathbf{I~,}%
\end{equation}
where $0\leq\tau\leq1$ is the transmissivity of the channel and $\bar{n}\geq0
$ its thermal number. A single-mode lossy channel can always be dilated into a
beam-splitter with transmissivity $\tau$, which mixes the incoming mode with
an environmental thermal mode with $\bar{n}$ mean photons.

\subsection{Symplectic spectrum\label{APP_subSPECTRUM}}

A central result in the theory of Gaussian states is Williamson's
theorem~\cite{Williamson,RMP}. Given a CM\ $\mathbf{V}$, there is a symplectic
matrix $\mathbf{S}$ realizing the diagonalization
\begin{equation}
\mathbf{V=SWS}^{T},~\mathbf{W=}\bigoplus\limits_{k=1}^{n}\nu_{k}\mathbf{I~,}%
\end{equation}
where the diagonal matrix $\mathbf{W}$\ is known as the Williamson normal
form, and $\{\nu_{1},\cdots,\nu_{n}\}$ are the $n$ symplectic eigenvalues of
the CM. Since $\det\mathbf{S}=1$, we have that%
\begin{equation}
\det\mathbf{V=}%
{\displaystyle\prod\limits_{k=1}^{n}}
\nu_{k}^{2}~.
\end{equation}

Using the symplectic spectrum, we can write the uncertainty principle of
Eq.~(\ref{unc_PRINC}) in the equivalent form%
\begin{equation}
\mathbf{V}>0,~\nu_{k}^{2}\geq1~, \label{BFnew}%
\end{equation}
which implies $\nu_{k}\geq1$~\cite{Note1}. The symplectic spectrum fully
determines the entropic and purity properties of the Gaussian states. In fact,
the von Neumann entropy $S(\rho):=-$Tr$(\rho\log\rho)$ of an arbitrary
$n$-mode Gaussian state is expressed by~\cite{RMP}%
\begin{equation}
S(\rho)=\sum_{k=1}^{n}h(\nu_{k})~, \label{VN_entropy}%
\end{equation}
where the function $h(x)$ is given in Eq.~(\ref{hVonNEUMANN}), with the
logarithmic base equal to $2$ for bits or \textquotedblleft$e$%
\textquotedblright\ for nats. The purity of a Gaussian state is given by%
\begin{equation}
\mu_{p}(\rho):=\mathrm{Tr}\rho^{2}=\frac{1}{\sqrt{\det\mathbf{V}}}=%
{\displaystyle\prod\limits_{k=1}^{n}}
\nu_{k}^{-1}~, \label{purity}%
\end{equation}
so that the state is pure (mixed) iff $\det\mathbf{V}=1$ ($>1$).

In our paper, we consider situations where the symplectic eigenvalues are
large. In this case, it is useful to use the asymptotic expansion%
\begin{equation}
h(x)\simeq\log\frac{e}{2}x+O\left(  \frac{1}{x}\right)  ~, \label{EXPA}%
\end{equation}
which is valid for large $x$. In particular, if the whole symplectic spectrum
is diverging ($\nu_{k}$ large for any $k$), then we can use Eq.~(\ref{EXPA})
to write the asymptotic formula%
\begin{equation}
S(\rho)\simeq\log\left(  \frac{e}{2}\right)  ^{n}\sqrt{\det\mathbf{V}}%
=\log\frac{1}{\mu_{p}(\rho)}+n\log\frac{e}{2}~, \label{connect}%
\end{equation}
where the entropy is directly related to the purity of the Gaussian state.

\subsection{Two-mode Gaussian states and bipartite
entanglement\label{APP_sub2modes}}

Since we are interested in the distribution of bipartite entanglement, we
devote a specific section to Gaussian states of two bosonic modes and their
separability properties. Let us consider two modes, $A$ and $B$, with
quadrature vector $\mathbf{\hat{x}}^{T}:=(\hat{q}_{A},\hat{p}_{A},\hat{q}%
_{B},\hat{p}_{B})$. We assume that these modes are described by a zero-mean
Gaussian state $\rho_{AB}$. Since $\mathbf{\bar{x}=0}$, this state is fully
characterized by its CM, that we can always express in the blockform
\begin{equation}
\mathbf{V}=\left(
\begin{array}
[c]{cc}%
\mathbf{A} & \mathbf{C}\\
\mathbf{C}^{T} & \mathbf{B}%
\end{array}
\right)  ~, \label{BlockFORM}%
\end{equation}
where $\mathbf{A}$, $\mathbf{B}$ and $\mathbf{C}$ are $2\times2$ matrices.
Finding the symplectic spectrum is straightforward, since~\cite{Sera, RMP}%
\begin{equation}
\nu_{\pm}=\sqrt{\frac{\Delta\pm\sqrt{\Delta^{2}-4\det\mathbf{V}}}{2}}~,
\label{symFORM}%
\end{equation}
where $\Delta:=\det\mathbf{A}+\det\mathbf{B}+2\det\mathbf{C}$. The uncertainty
principle is then equivalent to the bona-fide condition%
\begin{equation}
\mathbf{V}>0,~\nu_{-}^{2}\geq1~, \label{bonaFIDE}%
\end{equation}
where $\nu_{-}$ is the smallest symplectic eigenvalue.

As an example, consider the two-mode CM\ of the form
\begin{equation}
\mathbf{V}=\left(
\begin{array}
[c]{cccc}%
\omega &  & g & \\
& \omega &  & g^{\prime}\\
g &  & \omega & \\
& g^{\prime} &  & \omega
\end{array}
\right)  \label{CM_APP2}%
\end{equation}
where $\omega\geq1$ quantifies the thermal noise present in each mode, while
$g$ and $g^{\prime}$ are cross-correlation parameters. This is the CM of the
correlated-noise Gaussian environment which is defined in the main text [see
Eq.~(\ref{EVE_cmAPP})]. It is easy to derive simple bona-fide conditions in
terms of the parameters $\omega$, $g$ and $g^{\prime}$. In particular, given
$\omega\geq1$, we can easily derive conditions for the two correlation
parameters, by imposing Eq.~(\ref{bonaFIDE}). The positivity $\mathbf{V}>0$ is
equivalent to the positivity of the principal minors of the matrix. The
positivity of the first two minors is trivially implied by $\omega\geq1$. The
third minor gives
\begin{equation}
\det\left(
\begin{array}
[c]{ccc}%
\omega & 0 & g\\
0 & \omega & 0\\
g & 0 & \omega
\end{array}
\right)  >0\Leftrightarrow\omega(\omega^{2}-g^{2})>0,
\end{equation}
which is equivalent to%
\begin{equation}
|g|<\omega~. \label{gCON1}%
\end{equation}
The fourth minor corresponds to the determinant%
\begin{equation}
\det\mathbf{V}=(\omega^{2}-g^{2})(\omega^{2}-g^{\prime2})~, \label{detVapp}%
\end{equation}
and its positivity $\det\mathbf{V}>0$ leads to the condition%
\begin{equation}
|g^{\prime}|<\omega~. \label{gCON2}%
\end{equation}
We then derive the symplectic spectrum of the CM~(\ref{CM_APP2}), which is
given by
\begin{equation}
\nu_{\pm}=\sqrt{\omega^{2}+gg^{\prime}\pm\omega\left\vert g+g^{\prime
}\right\vert }~. \label{spectrum123}%
\end{equation}
Now, we find that $\nu_{-}^{2}\geq1$ is equivalent to~\cite{Note2}%
\begin{equation}
\omega^{2}+gg^{\prime}-1\geq\omega\left\vert g+g^{\prime}\right\vert ~.
\label{gCON3}%
\end{equation}
Thus, an environmental CM of the form~(\ref{CM_APP2}) with thermal noise
$\omega\geq1$ is a bona-fide CM if the correlation parameters $g$ and
$g^{\prime}$ satisfy the three bona-fide conditions of Eqs.~(\ref{gCON1}),
(\ref{gCON2}) and~(\ref{gCON3}) which are also given in
Eq.~(\ref{CMconstraints}) of the main text. Using Eqs.~(\ref{purity})
and~(\ref{detVapp}), we also see that the purity of the Gaussian state with
CM~(\ref{CM_APP2}) is given by
\begin{equation}
\mu_{p}=\left[  (\omega^{2}-g^{2})(\omega^{2}-g^{\prime2})\right]  ^{-1/2}.
\label{purCON}%
\end{equation}
Its von Neumann entropy is computed by using Eq.~(\ref{VN_entropy}) and the
spectrum in Eq.~(\ref{spectrum123}).

It is easy to study the separability properties of an arbitrary two-mode
Gaussian state with CM in the generic blockform of Eq.~(\ref{BlockFORM}). In
fact, it is sufficient to compute the smallest partially-transposed symplectic
(PTS) eigenvalue $\varepsilon$. This eigenvalue can be computed using the
formula of $\nu_{-}$, given in Eq.~(\ref{symFORM}), proviso that we replace
$\Delta$ with $\tilde{\Delta}=\det\mathbf{A}+\det\mathbf{B}-2\det\mathbf{C}$.
Then, a Gaussian state is separable (entangled) if and only if $\varepsilon
\geq1$ ($\varepsilon<1$). More strongly, this eigenvalue provides a
quantification of entanglement, being an entanglement monotone for two-mode
Gaussian states. In fact, it is monotonically related to the log-negativity
$\mathcal{N}=\max\{0,-\log\varepsilon\}$, which is another entanglement
monotone and can be expressed in entanglement bits (ebits).

The log-negativity $\mathcal{N}$ provides an upper-bound to the distillable
entanglement, corresponding to the average number of ebits per copy which can
be extracted from infinitely-many copies of the state $\rho_{AB}\otimes
\rho_{AB}\otimes\ldots$ We can also consider a lower-bound to distillable
entanglement. According to the hashing inequality~\cite{Deve,Deve2}, this
lower bound is given by the coherent information~\cite{CohINFO,CohINFO2}
\begin{equation}
I(A\rangle B)=S(\rho_{B})-S(\rho_{AB})~,
\end{equation}
where $\rho_{B}=\mathrm{Tr}_{A}(\rho_{AB})$ is the reduced state of Bob, here
corresponding to a zero-mean Gaussian state with CM $\mathbf{B}$. Using
Eq.~(\ref{VN_entropy}), the coherent information becomes%
\begin{equation}
I(A\rangle B)=h(\nu_{B})-h(\nu_{-})-h(\nu_{+})~,
\end{equation}
where $\nu_{B}=\sqrt{\det\mathbf{B}}$ is the symplectic eigenvalue
of\ $\mathbf{B}$, and $\{\nu_{-},\nu_{+}\}$ is the symplectic spectrum of
$\mathbf{V}$. In the case where these eigenvalues are all large, we can use
the expansion of Eq.~(\ref{EXPA}) to get the formula%
\begin{equation}
I(A\rangle B)\simeq\log\frac{2}{e}\sqrt{\frac{\det\mathbf{B}}{\det\mathbf{V}}%
}~. \label{I_div2}%
\end{equation}

More precisely, the coherent information is a lower bound to the distillable
entanglement which is achievable by means of one-way protocols between Alice
and Bob (one-way distillability). In these protocols, Alice applies a quantum
instrument to her copies, i.e., a collection of completely-positive
trace-preserving maps labelled by a classical index $k$. Then, she classically
communicates $k$ to Bob, who applies a conditional quantum operation on his
copies (see Ref.~\cite{Deve}\ for more details).

\subsection{EPR\ correlations\label{APP_subEPR}}

The most important example of two-mode Gaussian state is the two-mode squeezed
vacuum state, also known as the EPR state. This state represents the most
common source of continuous-variable entanglement.

An EPR state has zero mean and CM given by Eq.~(\ref{EPRcmMU}) in the main
text. The quadratures $\mathbf{\hat{x}}^{T}:=(\hat{q}_{A},\hat{p}_{A},\hat
{q}_{B},\hat{p}_{B})$ have equal variances
\begin{equation}
\left\langle \hat{q}_{A}^{2}\right\rangle =\ldots=\left\langle \hat{p}_{B}%
^{2}\right\rangle =\mu\geq1~,
\end{equation}
and maximal correlations
\begin{equation}
\left\langle \hat{q}_{A}\hat{q}_{B}\right\rangle =-\left\langle \hat{p}%
_{A}\hat{p}_{B}\right\rangle =\mu^{\prime}:=\sqrt{\mu^{2}-1}~.
\end{equation}
The parameter $\mu$\ can be used to quantify the entanglement between the two
modes $A$ and $B$. In fact, the smallest PTS\ eigenvalue takes the form%
\begin{equation}
\varepsilon=\sqrt{2\mu\left(  \mu-\mu^{\prime}\right)  -1}~,
\end{equation}
which is monotonically decreasing in $\mu$. We can easily check that we have
entanglement ($\varepsilon<1$) for any $\mu>1$.

The EPR\ correlations can be characterized in terms of the variance of the EPR
quadratures
\begin{equation}
\hat{q}_{-}:=\frac{\hat{q}_{A}-\hat{q}_{B}}{\sqrt{2}}~,~\hat{p}_{+}%
:=\frac{\hat{p}_{A}+\hat{p}_{B}}{\sqrt{2}}~. \label{EPRquad}%
\end{equation}
The EPR\ condition corresponds to $V(\hat{q}_{-})=V(\hat{p}_{+})<1$, which
means that the quadrature correlations between the two modes are below the
vacuum noise. For the EPR\ state described by Eq.~(\ref{EPRcmMU}) indeed we
have
\begin{equation}
V(\hat{q}_{-})=V(\hat{p}_{+})=\mu-\mu^{\prime}~,
\end{equation}
which is less than $1$ for any $\mu>1$. For this type of state,
EPR\ correlations and entanglement are equivalent conditions. In general, for
an arbitrary two-mode Gaussian state, the presence of EPR\ correlations is
only a sufficient condition for entanglement.

In the limit of large entanglement ($\mu\rightarrow\infty$), the EPR\ state
becomes ideal. In fact, we have $V(\hat{q}_{-})=V(\hat{p}_{+})\rightarrow0$,
which corresponds to realizing the ideal EPR\ correlations $\hat{q}_{A}%
=\hat{q}_{B}$ and $\hat{p}_{A}=-\hat{p}_{B}$, where positions (momenta) are
perfectly correlated (anticorrelated). Note that an alternative EPR\ state can
be defined using the other reflection matrix $-\mathbf{Z}$ in the place of
$\mathbf{Z}$ in Eq.~(\ref{EPRcmMU}). This alternative state has EPR
correlations in the quadratures $\hat{q}_{+}$ and $\hat{p}_{-}$, therefore
tending to realize the ideal conditions $\hat{q}_{A}=-\hat{q}_{B}$ and
$\hat{p}_{A}=\hat{p}_{B}$. In our work, we do not consider this state, but
similar results can easily be derived.

\subsection{Quantum discord and Gaussian discord\label{APP_subDISCORD}}

Quantum entanglement is synonymous of quantum correlations for pure states.
For mixed states the scenario is more subtle. In fact, separable mixed states
may still contain features of quantumness, one of which is known as quantum
discord~\cite{Qdiscord,Qdiscord2,RMPdis}. Quantum discord derives from
different quantum extensions of the notion of classical mutual information.

Classically, the mutual information of two random classical variables, $A$ and
$B$, is defined as
\begin{equation}
I(A:B)=H(A)+H(B)-H(A,B)~, \label{qI1}%
\end{equation}
where $H$ is the Shannon entropy. Using the chain rule for the entropy, we can
also write the equivalent formulation%
\begin{equation}
I(A:B)=H(A)-H(A|B)~, \label{qI2}%
\end{equation}
where $H(..|..)$ is the conditional Shannon entropy.

In quantum mechanics, we can have two inequivalent generalizations, the first
being provided by the quantum mutual information which accounts for the total
correlations present in a bipartite state $\rho_{AB}$. It is expressed in
terms of the von Neumann entropy as
\begin{equation}
I(\rho_{AB})=S(\rho_{A})+S(\rho_{B})-S(\rho_{AB})~.
\end{equation}
Another generalization is given by the entropic quantity%
\begin{equation}
C(\rho_{AB})=S(\rho_{A})-\inf_{\mathcal{M}_{B}}S(A|\mathcal{M}_{B})~,
\label{CDEVE}%
\end{equation}
where $\mathcal{M}_{B}=\{M_{k}\}$ is a POVM acting on system $B$ and
\begin{equation}
S(A|\mathcal{M}_{B}):=\sum_{k}p_{k}S(\rho_{A|k})~,
\end{equation}
where $p_{k}$ is the probability of the outcome $k$ and $\rho_{A|k}%
:=\mathrm{Tr}_{B}(\rho_{AB}M_{k})$ is the conditional state of $A$.

The quantity $C(\rho_{AB})$, which is expressed in bits or classical bits
(cbits), quantifies the classical correlations within the state, corresponding
to the maximal amount of common randomness which can be extracted using local
operations and one-way classical communication~\cite{DEVEcommon}.

Quantum discord aims to quantify the quantum correlations in the state. Thus,
it is defined as the difference between total and classical correlations%
\begin{equation}
D(\rho_{AB})=I(\rho_{AB})-C(\rho_{AB})~, \label{DISCORDeq}%
\end{equation}
and measured in bits or discordant bits (dbits). It is important to note that
Eq. (\ref{CDEVE}) is generally non-symmetric under system permutation. As a
result, we have that the $B$-type quantities $C(\rho_{BA})$ and $D(\rho_{BA})$
are generally different from the $A$-type quantities of Eqs.~(\ref{CDEVE})
and~(\ref{DISCORDeq}). Such ambiguity clearly disappears for symmetric states
($\rho_{AB}=\rho_{BA}$), as the states considered in our paper.

For bosonic states, one can define the Gaussian discord by restricting the
previous minimization from arbitrary to Gaussian POVMs~\cite{GerryD,ParisD}.
It is clear that Gaussian discord represents an upper bound to the actual
discord, and the classical correlations of Eq.~(\ref{CDEVE}) restricted to
Gaussian POVMs represent a lower bound to the actual classical correlations of
the state. Gaussian discord is conjectured to be the actual discord in the
case of bosonic Gaussian states, for which we know closed
formulas~\cite{GerryD,ParisD}. In fact, consider a two-mode Gaussian state
with CM in the blockform of Eq.~(\ref{BlockFORM}) and symplectic spectrum
$\{\nu_{-},\nu_{+}\}$. Then, the (restricted) classical correlations and
Gaussian discord are given by%
\begin{align}
C(\rho_{AB})  &  =h\left(  \sqrt{\det\mathbf{A}}\right)  -\Sigma
,\label{Cgauss}\\
D(\rho_{AB})  &  =h\left(  \sqrt{\det\mathbf{B}}\right)  -h(\nu_{-})-h(\nu
_{+})+\Sigma, \label{Dgauss}%
\end{align}
where the expression of the $\Sigma$-term is given in Ref.~\cite{GerryD}.
Apart from the specific case of tensor-product states $\rho_{AB}=\rho
_{A}\otimes\rho_{B}$, all two-mode Gaussian states have non-zero Gaussian
discord~\cite{GerryD} and non-zero (actual) discord~\cite{VerDIS}. In
particular, Gaussian discord $D>1$ implies the presence of
entanglement~\cite{GerryD}.


\begin{thebibliography}{99}
\bibitem {NOTAdeco}In general, a decoherence-free subspace is a subspace of
the state-space which evolves unitarily under the action of the quantum
channel. This subspace is invariant when no evolution occurs.
\bibitem {NoteCM1}In fact, the global input state $\rho_{AB}\otimes\rho
_{E_{2}}$ has CM $\mathbf{V}_{AB}(\mu)\oplus\omega\mathbf{I}$. This CM is
subject to the symplectic transformation $\mathbf{I}\oplus\mathbf{S}(\tau)$,
where $\mathbf{S}(\tau)$ is the beam splitter matrix of Eq.~(\ref{BSmatrix})
acting on modes $B$ and $E_{2}$. Tracing out the environment from the output
CM, we get Eq.~(\ref{ABpCM}).
\bibitem {Notation}Minor remark on matrix notation. We use $\rho_{ijkl}$ when
we refer to kets $\left\vert i,j\right\rangle $ and bras $\left\langle
k,l\right\vert $. We prefer to use $\rho_{ij}^{kl}$ when we distinguish
between system $A$ ($\left\vert i\right\rangle _{A}\left\langle j\right\vert
$) and system $B$ ($\left\vert k\right\rangle _{B}\left\langle l\right\vert
$). Clearly, we have $\rho_{ijkl}=\rho_{ik}^{jl}$.
\end{thebibliography}
\end{document}